\documentclass[twocolumn]{aastex6} 
\usepackage{boxedminipage}
\usepackage{amsmath,amsfonts,amssymb}
\usepackage{comment}
\usepackage{natbib}
\usepackage{tablefootnote}
\usepackage{threeparttable}
\usepackage{multirow}
\usepackage{CJKutf8} 
\usepackage{cancel}
\usepackage{afterpage}
\usepackage{gensymb}
\usepackage{tabularx}
\begin{document}

\title{Precision Time-series Photometry in the Thermal Infrared with a\\ `Wall-Eyed' Pointing Mode at the Large Binocular Telescope}

\author{Eckhart Spalding\altaffilmark{1}, Phil Hinz\altaffilmark{1}, Andrew Skemer\altaffilmark{4}, John Hill\altaffilmark{2}, Vanessa P. Bailey\altaffilmark{3}, \and
Amali Vaz\altaffilmark{1}}

\altaffiltext{1}{Steward Observatory, University of Arizona, 933 North Cherry Ave. Tucson, AZ 85721, USA}
\altaffiltext{2}{University of Arizona, Large Binocular Telescope Observatory, 933 North Cherry Ave., Tucson, AZ 85721, USA}
\altaffiltext{3}{Kavli Institute for Particle Astrophysics and Cosmology, Stanford University, Stanford, CA 94305, USA}
\altaffiltext{4}{Department of Astronomy and Astrophysics, University of California, Santa Cruz, 95064, USA}

\begin{abstract}

Time-series photometry taken from ground-based facilities is improved with the use of comparison stars due to the short timescales of atmospheric-induced variability.  However, the sky is bright in the thermal infrared (3-5 $\mu$m), and the correspondingly small fields-of-view of available detectors make it highly unusual to have a calibration star in the same field as a science target.  Here we present a new method of obtaining differential photometry by simultaneously imaging a science target and a calibrator star, separated by $\lesssim$2 amin, onto a 10$\times$10 asec$^{2}$ field-of-view detector.  We do this by taking advantage of the LBT's unique binocular design to point the two co-mounted telescopes apart and simultaneously obtain both targets in three sets of observations. Results indicate that the achievable scatter in $L_{S}$-band ($\lambda_{c}=3.3$ $\mu$m) is at the percent level for bright targets, and possibly better with heavier sampling and characterization of the systematics.

\end{abstract}


\section{Introduction} \label{sec:intro}

Photometry in the thermal infrared (3-5 $\mu$m) has a number of potential applications associated with exoplanetary systems at various stages of their evolution. The thermal infrared probes circumstellar disk material at hundreds of Kelvin in the inner $\lesssim$10 AU of the disk, and water ice has a wide rovibrational transition band at $\sim$3 $\mu$m that can provide constraints on debris disk dust composition \citep{henning2013chemistry,rodigas2014does}. Infrared emission can also be a signature of stellar accretion and mass loss \citep{polsdofer2015examining}. 

Rapid-cadence, time-series photometry is particularly valuable for placing constraints on the composition and vertical structure of brown dwarf and exoplanet atmospheres. This is done by observing flux variations on the timescales of the rotation periods of brown dwarfs, or the passage of an exoplanet in front of (or behind) its host star \citep{winn2010exoplanet,buenzli2012vertical}. For example, the $L$- and $L_{S}$-bands can allow the characterization of a methane bandhead in the atmospheres of either brown dwarfs or giant planets, and can help infer the presence of disequilibrium chemistry \citep{skemer2012first,skemer2014directly}. 

\begin{figure}
\centering
\includegraphics[width=1.0\linewidth]{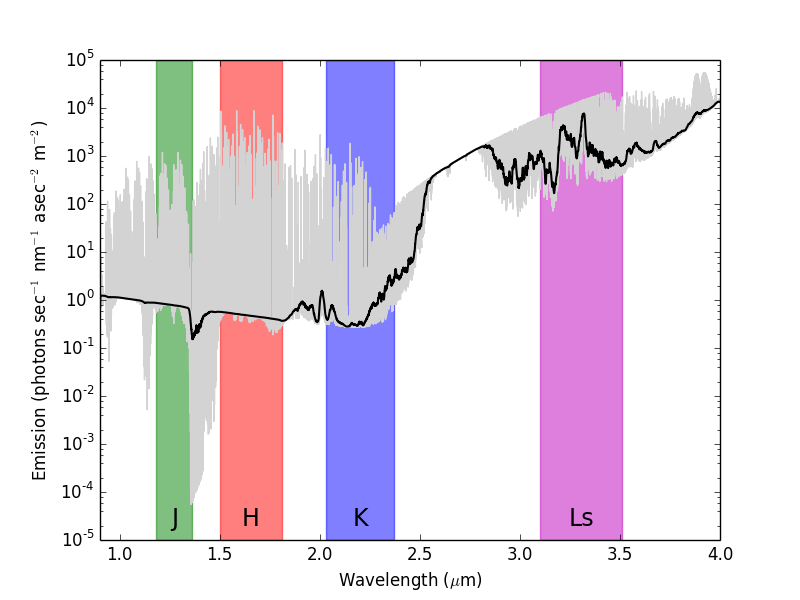}
\caption{A plot of sky emission above Mauna Kea. As wavelengths increase beyond $K$-band, the blackbody of the atmosphere becomes the dominant component of the sky brightness. Narrow molecular airglow lines are visible $<$2.5 $\mu$m. (Data from Gemini Observatory. Emission resolution is 2$\times10^{-2}$ nm (grey), with a smoothing of 20 nm (black). Reprinted from \citet{spalding2016infrared}.)} 
\label{fig:maunakeaemission}
\end{figure}

Attempts to make precise measurements at thermal infrared wavelengths face a number of obstacles. Space-based facilities are currently limited to the `warm' Spitzer mission's 3.6 and 4.5 $\mu$m broadband channels, which are insufficient to overcome some degeneracies in atmospheric models \citep{kammer2015spitzer}. Ground-based observations are plagued by systematic effects due to the Earth's atmosphere (e.g., \citet{croll2015near}). Time-series photometry in the thermal infrared is particularly challenging to obtain from the ground due to two interlocking factors: the brightness and variability of the combined backgrounds of the telescope and atmosphere, and the small fields-of-view of detectors sensitive to these wavelengths. Fig. \ref{fig:maunakeaemission} shows how the brightness of the atmosphere alone increases by a few decades in the thermal infrared as blackbody emission becomes dominant. 

To counter the rapid saturation of detector pixels, fields-of-view of thermal infrared detectors ($\sim$few 10 asec across) tend to be restricted so as to spread the photons from the background more thinly across the array. In addition, plate scales of detectors meant for adaptive optics (AO)-corrected imaging are made so as to finely sample the point spread function (PSF). AO acts to minimize the sky footprint beneath the PSF and thereby minimize the sky component of the photon noise in strongly background-limited observations. This means that even if the substantial background effects can be decorrelated from the science signal by utilizing a bright comparison star, the tiny fields-of-view conspire to make it unlikely that bright comparison stars will be available to perform differential photometry. This quandary is distinct from the need to find a sufficiently bright star to generate the AO correction in the first place. (Indeed, it is frequently the case in AO-corrected observations of protoplanetary or exoplanetary systems that the host star itself is bright enough for AO correction.)

One option for obtaining precise thermal infrared photometry is to slew the telescope back and forth between the science target and a comparison star in order to remove thermal atmospheric variations (e.g., \citet{stephens2001band, deming20073}). A disadvantage of this strategy is that it incurs a loss of observing efficiency and fails to capture correlated variability on timescales shorter than the integration or slew time. An alternative strategy is to attempt to remove atmospheric effects entirely in the post-processing phase without a comparison star, but getting reliable results has been challenging \citep{mandell2011non}. Yet another alternative is to couple ground-based spectroscopy with correlation techniques, which can potentially be used to detect relative signal depths of $\sim 10^{-3}$ or possibly even smaller, comparable with space-based precisions \citep{birkby2013detection,zellem2014ground}. However, spreading photons too thinly among spectral bins can lead to low signal-to-noise. High-resolution spectroscopy is also geared more towards sharp absorption features in a planet's atmosphere, rather than continuum.

\begin{figure}
\centering
\includegraphics[width=1.0\linewidth]{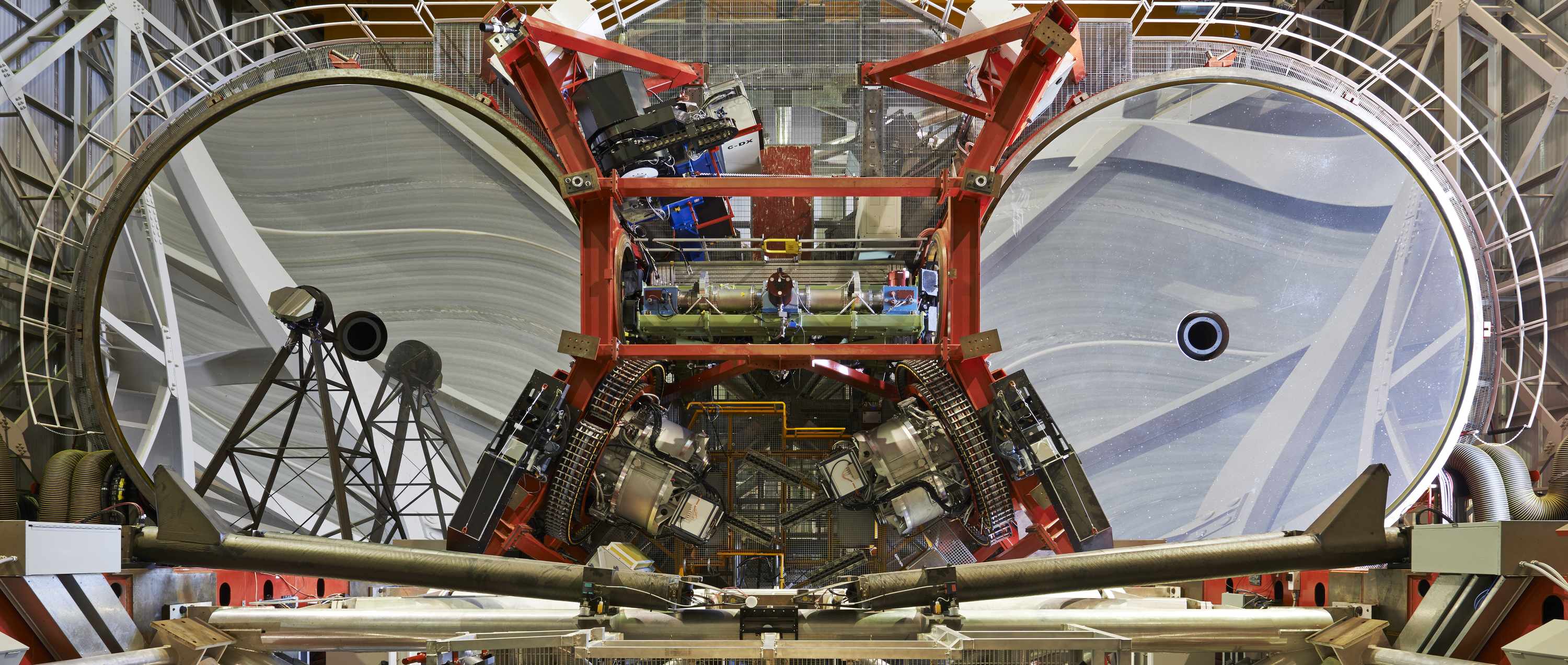}
\caption{The twin primary mirrors of the LBT, with one of the tertiary mirror swing arms in place. The LBTI instrument combines the two beams between the mirrors. (Image credit: LBTO, Enrico Sacchetti; with permission from LBTO. Reprinted from \citet{spalding2016infrared}.)} 
\label{fig:lbtPrimaries}
\end{figure}

The Large Binocular Telescope (LBT) on Mt. Graham, Arizona, is a large, 2$\times$8.4-m  telescope with adaptive optics and a unique binocular design that can provide an alternative technique for performing thermal infrared differential photometry on accessible targets.\footnote{At the LBT's latitude, targets with $-15\degree< DEC < +80\degree$ have minimum airmasses $\leq1.5$.} The LBT design is based on twin Gregorian telescopes sitting on a common mount with a center-to-center separation of 14.4 m \citep{hill2010large}. Fig. \ref{fig:lbtPrimaries} shows the primary mirrors on either side of instruments mounted at bent Gregorian foci. The primary and secondary mirrors sit on hexapods which can be translated and rotated so as to point the telescopes apart by up to $\sim$2 amin of separation \citep{rakich2011maintaining}. In such an event, the optics of each telescope remain collimated and do not induce additional optical aberrations on the beam. 

The LBT has an AO system which minimizes warm optics in the beam path by using an adaptive secondary mirror in each telescope, each of which has 672 actuators \citep{esposito2010first}. Spatial sampling of the wavefront is set by the binning of the wavefront sensor detector. Up to 400/153/66/36 spatial modes can be corrected in 1/2/3/4 binning of wavefront subapertures. (Thirty unbinned subapertures can be fit across the pupil diameter.) Fainter stars use higher binning so as to gather enough photons to reconstruct the aberrated wavefront, and the Strehl ratio decreases with decreasing number of modes.

The LBTI instrument is mounted at a pair of bent Gregorian foci, where a beam combiner parallelizes the two beams and directs them into the science instrument \citep{hinz2004large,hinz2008nic}. The science instrument has two infrared cameras: LMIRcam (sensitive to 1.5-5 $\mu$m) \citep{leisenring2012sky} and NOMIC (currently 7-14 $\mu$m) \citep{hoffmann2014operation}. 

In this paper, we present observations carried out with the LMIRcam detector in $L_{S}$-band to quantify the achievable photometric precision with the LBT's `wall-eyed' mode. We observed three sets of targets in varying environmental conditions, as described in Section \ref{sec:observations}. The analysis is described in Section \ref{sec:analysis}, which includes a description of the optimal photometry parameters in subsection \ref{subsec:minRedNoise}, the search for contributors to the systematics in subsection \ref{subsec:systematics}, and an exploration of parametric systematics models in subsection \ref{subsec:fitting}. The results are described in Section \ref{sec:results}. We discuss the results and systematics in Section \ref{sec:discussion} and summarize in Section \ref{sec:summary}.

\section{Observations} \label{sec:observations}

\subsection{General Strategy and Telemetry} \label{subsec:lbti}

\begin{figure}
\centering
\includegraphics[width=1.0\linewidth]{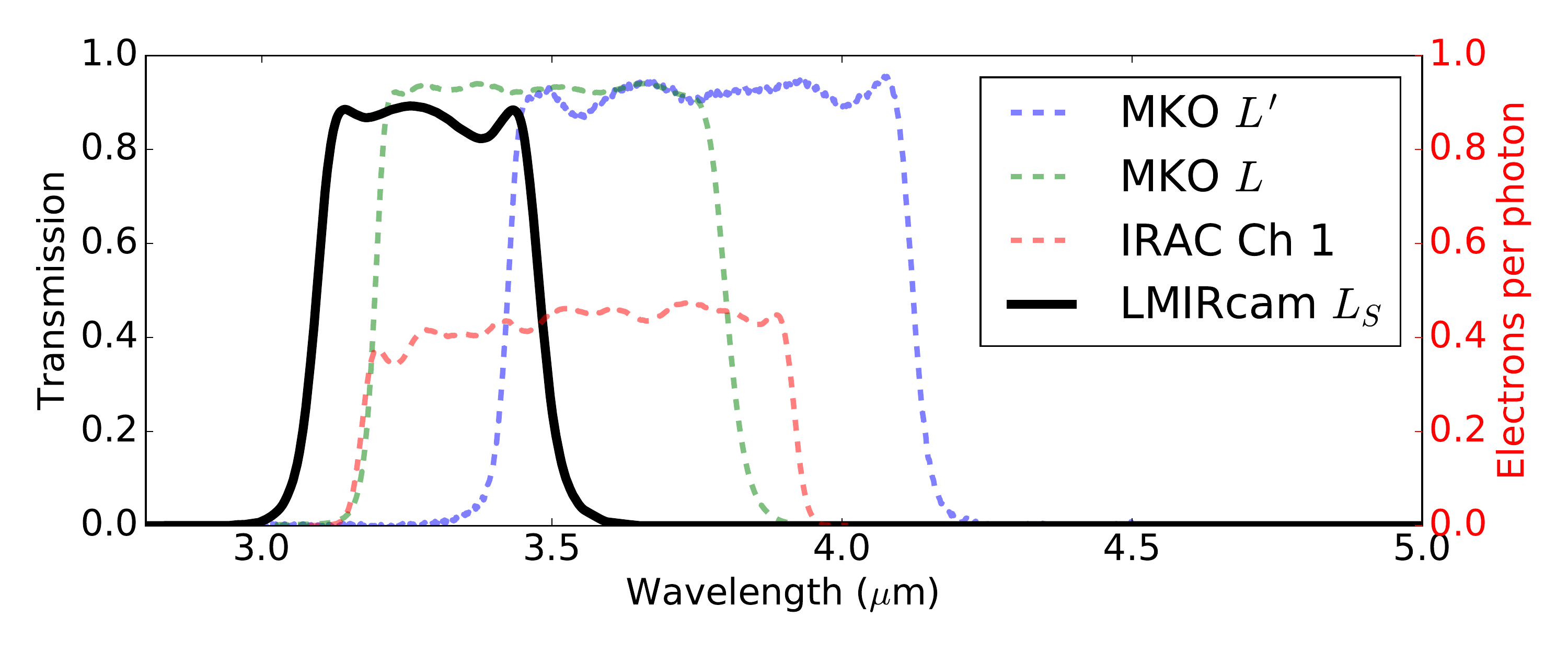}
\caption{Comparative transmission curves for the Mauna Kea Observatories (MKO) $L$ and $L'$ filters \citep{simons2002mauna,tokunaga2002mauna}, the average transmission of the Spitzer/IRAC ``3.6 $\mu$m'' Channel 1 \citep{quijadaa2004angle}, and the LMIRcam JDSU $L_{S}$ filter. Units of the IRAC Channel 1 should be read using the units along the right-hand y-axis, and the other curves using the left-hand axis.} 
\label{fig:filters}
\end{figure}

We observed three targets using the LMIRcam detector: a target with constant flux, a primary transit of the exoplanet XO-2Nb in front of its host star, and a secondary eclipse of exoplanet HD 189733b (Table \ref{table:observing_log}). Each observation also included a comparison star. In each case, the beams from the two apertures were moved to place the target and the comparison star PSFs, along with the two overlapping skies, onto the LMIRcam detector. We used a $L_{S}$ filter made by the company JDS Uniphase (JDSU) (Fig. \ref{fig:filters}). This filter transmits with width 0.4 $\mu$m at a central wavelength of $\lambda_{c}=3.3$ $\mu$m, shortward of the central wavelengths of the Mauna Kea Observatories (MKO) $L$-bands and Spitzer Channel 1. 

\floattable
\begin{deluxetable}{l c | ccccc | ccccc | ccc}
\rotate
\tablewidth{200pt}
\tabletypesize{\scriptsize}
\tablecaption{Observing log\label{tab:observing_log}}
\tablehead{
\colhead{Dataset} & \colhead{UT Date} & \colhead{Left Target} & \colhead{$m_{K}$} &  \colhead{$\nu_{AO}$ (Hz)$^{a}$} & \colhead{Bin$^{b}$} & \colhead{$r_{c}$ ('')$^{c}$} & \colhead{Right Target} & \colhead{$m_{K}$} &  \colhead{$\nu_{AO}$ (Hz)$^{a}$} & \colhead{Bin$^{b}$} & \colhead{$r_{c}$ ('')$^{c}$}& \colhead{$\Delta\theta$ ('')$^{d}$} & \colhead{Duration (hr)$^{e}$} & \colhead{Eff$^{f}$} }
\startdata
Constant targets & 2014 Oct 2& 	ADS 1571 B & 7.3	& 990 & 2 & 0.61 &	HR 567 & 5.2 	& 990 & 1 & 1.22 &		42.6 & 1.49 & 0.41 \\
Primary transit & 2015 Feb 8& 		XO-2S & 9.3 & 	 990 & 3	& 0.41	&	XO-2N* & 9.3  & 990 & 3	&	0.41 &	31.2 & 5.13 & 0.83 \\
Secondary eclipse & 2015 Jun 13& 	HD 189733* & 5.5 	& 990 & 1 & 1.22&	2MJ20004297+2242342 & 9.3  & 748 & 4 & 0.30 & 11.4 & 2.74 & 0.58 \\
\enddata
\tablenotetext{*}{Science target}
\tablenotetext{a}{The AO loop frequency.}
\tablenotetext{b}{Wavefront sensor subaperture binning size. This is the number $N$ by which to bin $N\times N$ base subapertures across a pupil 30 base subapertures in diameter.}
\tablenotetext{c}{The AO control radius, within which the AO correction is active. This depends on the number of deformable modes applied, and is $r_{c}\approx3.3 \mu m/(2d)$, where $d$ is the intersubaperture spacing.}
\tablenotetext{d}{On-sky angular separation between the two targets. Based on models of turbulence, at these wavelengths the isoplanatic angle within which an AO correction for one target is valid is on the order of $\sim$10 asec at zenith. This decreases by an order of magnitude at an elevation of $30^{\circ}$ \citep{tyson2004field}.}
\tablenotetext{e}{The duration is the total length of time, or `wall-clock time', from the first to the last integration. This includes overheads between integrations, but does not include common AO overheads of several minutes prior to the first integration so as to acquire the targets and make the initial closure of the AO loops.}
\tablenotetext{f}{The observing efficiency ratio, which is found by dividing the total target integration time by the total duration of the observation.}
\label{table:observing_log}
\end{deluxetable}

The AO-corrected PSFs were heavily oversampled at a plate scale of $10.7$ masec per pixel, which we determined using 101 stellar pair baselines in the Trapezium cluster, and which is consistent with the plate scale found by \citet{maire2015leech}. The first dark Airy ring of a diffraction-limited, monochromatic 3.3 $\mu$m PSF corresponds to a radius of 99 masec and encloses a total of $\sim$270 pixels. The AO correction involved application of orthogonal Karhunen–Lo\`{e}ve (KL) modes to the deformable secondary mirror at a rate of up to 990 Hz. The greater the number of modes applied, the more effectively the deformable mirror can cancel out the wavefront deformation. Up to 400 KL modes were applied, though sometimes fewer modes were used depending on the size of the subaperture binning. 

The PSFs were initially placed in the top half of the detector array, and then offset up and down in elevation by $\sim$4.5 asec. LBTI does not have a derotator. As the objects rotated on the sky with the parallactic angle, the AO system kept the targets in place on the detector. As a result, the two targets individually rotated in place on the detector, but did not rotate around each other.

Concurrent to our observations, we gathered environmental telemetry including seeing, precipitable water vapor, and windspeed and wind direction as measured from the LBT roof. Seeing was measured by a Differential Image Motion Monitor (DIMM) attached to the telescope mount. The DIMM was pointed at a single bright star in the general direction of the telescope's pointing, and determined the seeing through the comparative motion of the star through two apertures. The seeing values in this work correspond to the general area of the observed object on the sky, and are not normalized to be along the zenith.

Water vapor is both highly absorptive and emissive in the infrared, and strongly affects atmospheric transmissivity and the level of the sky background. The amount of atmospheric water vapor is quantified as precipitable water vapor, or the height of the entire vertical column of water in the atmosphere if the water were condensed into liquid form. Precipitable water vapor was measured at the Heinrich Hertz Submillimeter Telescope $\sim$180 m away from the LBT, using a radiometer which is repeatedly tipped to measure the sky brightness temperature as a function of zenith angle, and fits a functional form for the sky brightness temperature to solve for the opacity $\tau$ at 225 GHz \citep{liutipper,chamberlin1994225}. A conversion factor of 20 converts $\tau$ into mm H$_{2}$O \citep{thomas2007calibration}.\footnote{\url{http://aro.as.arizona.edu/weather\_stats/standard\_tau\_plots.htm}}

\subsection{Constant Targets} \label{subsec:constantObservation}

The baseline observation of UT 2014 Oct 2 involved the two bright stellar targets ADS 1571 B and HR 567. (See Table \ref{tab:observing_log}.) Both targets were observed with separate apertures. Integrations were taken at 58 msec, which were then summed into 12-coadd arrays for a total integration time of 0.70 seconds per frame. Each nod position lasted 4.1 minutes and included 240 frames. Flats and darks were taken at the end of the night with the same integration time and number of co-adds. 

During this observation, seeing ranged between 0.9 and 1.35 asec, until a sudden $\sim$1.8 asec seeing burst near the end of the observation. Airmass changed very little, ranging between 1.18 and 1.19. Windspeeds ranged between $\sim$2 to 11 m/s. Precipitable water vapor went unrecorded because of equipment malfunction.

\subsection{XO-2Nb Primary Transit} \label{subsec:primaryObservation}

The XO-2 binary system is composed of the G9V type stellar members XO-2N and XO-2S. Given the Gaia DR1 parallax of XO-2S of $6.68\pm0.23$ masec \citep{lindegren2016gaia,prusti2016gaia,brown2016gaia}, this system is $150\pm5$ pc distant. The two stars have an angular separation of 31.2 asec, or a sky-projected separation of $\sim4.7\times10^{3}$ AU. Both stars are known to host planets, making the XO-2 system an observational laboratory for the evolution of planetary systems around binary main sequence hosts. The 0.6$M_{J}$ planet XO-2Nb transits in front of its host star with a period of 2.616 days \citep{burke2007xo,fernandez2009transit}, which has permitted atmospheric characterization in the optical \citep{sing2011gran,crouzet2012transmission,sing2012gtc}. 

We observed the XO-2 system on UT 2015 Feb 8 during a primary transit of XO-2Nb, with the aim of determining whether we could detect a larger transit depth due to an increased scale height of its atmosphere in $L_{S}$ through possible methane absorption. Environmental conditions were fair, with seeing below 1.2 asec except for a few short bursts. Precipitable water vapor levels remained below 4 mm H$_{2}$O. The airmass ranged between 1.05 and 1.43, and winds between $\sim$1 and 9 m/s. This dataset involved the highest airmass of the three datasets, and the maximum change in airmass between adjacent nods was $\sim10^{-5}$. Integration times were 0.99 sec, readouts were single co-adds, and every nod position included 200 frames and lasted 3.4 minutes. Flats and darks were taken immediately before and after the observation. At the end of the observing night, we also took additional flat and dark frames at integration times staggered by 29 msec, so as to provide linearity response information.

\subsection{HD 189733b Secondary Eclipse} \label{subsec:secondaryObservation}

HD 189733b is a 1.2$M_{J}$ planet orbiting a bright host star ($m_{K}=5.5$) \citep{cutri2003vizier,de2013detection} and is one of the nearest transiting or eclipsing exoplanet systems known. The Gaia DR1 parallax of HD 189733 is $50.40\pm0.22$ masec, equivalent to a distance of $19.84\pm 0.09$ pc \citep{lindegren2016gaia,prusti2016gaia,brown2016gaia}. Accordingly, HD 189733b has been the subject of numerous observations over the years to characterize its atmosphere (e.g., \citet{grillmair2008strong,pont2008detection,redfield2008sodium,desert2009search,sing2009transit,lee2012optimal,birkby2013detection,de2013detection,rodler2013detection,mccullough2014water,crouzet2014water,swain2014detection}). However, there has been some conflict among different claims of detections and non-detections of atmospheric species. Thus this system is well-positioned for photometric atmospheric characterization, but also illustrates the importance of obtaining independent measurements.

We attempted an observation of the secondary eclipse of HD 189733b behind its host star on UT 2015 June 13. As an occulted planet passes behind the host star, the light which is reflected and emitted by the planet are subtracted from the net flux. The relative flux decrement of the secondary eclipse of HD 189733b is 0.26\% as seen by Spitzer's 3.6 $\mu$m channel \citep{charbonneau2008broadband}. With sufficient precision, it would be possible to provide independent tests of previous claims of water absorption in the dayside spectrum of HD 189733b.

Integrations were taken for 0.23 sec each, and were collapsed into arrays of 5 co-adds of 1.16 sec total integration. Every nod included 200 frames and lasted 5.0 minutes. Seeing remained between 0.7 and 1.1 asec, except for a brief 1.3 asec burst. Airmass ranged between 1.02 and 1.15, and windspeed was $\sim$1 to 5 m/s. Precipitable water vapor levels were particularly high, varying rapidly on a timescale of a few minutes between 5.6 and 12.6 mm H$_{2}$O. Flats and darks were again taken immediately before and after the observation. We obtained data for a linearity correction at the end of the night. 

The right adaptive secondary mirror was only able to produce a mediocre correction of the comparison star, basing its correction on 36 orthogonal KL modes and the maximum wavefront sensor detector binning. In principle it should not be a problem that the left and right PSF morphologies were different. However, some minutes before the planet was expected to emerge from behind the host star, stresses in the right secondary mirror triggered it to rest in a safe mode. This occurred at a high elevation of $79.5^{\circ}$ a few minutes before the telescope reached the maximum elevation of the observation ($80.0^{\circ}$). The maximum elevation marks the time when the parallactic angle is changing most rapidly and the mirror mount must accordingly swing around rapidly to remain on-target. It took $\sim$8 minutes to recover the mirror from its safe state, after which the right PSF morphology had changed.

This safe state event was likely due to an equipment problem, since the guide star on the right side was on-axis and sufficiently bright for the AO loop to close, and there was no burst of wind at the time of the event. Safe state events typically occurred a few times every night. Beginning in 2016, however, software improvements have decreased the frequency of such events.

\section{Analysis} \label{sec:analysis}

\subsection{Data Reduction} \label{subsec:relative}

IDL routines were written in order to correct bad pixels, bias- and dark-subtract the arrays, linearize the individual pixel responses (except in the case of the constant target data, for which we did not have concurrent linearization data), flat-field the arrays, and background-subtract the results. 

During the reduction process it was realized that the constant target dataset had inadequate sky flats. In that dataset, the flat integration times of 58 msec were chosen to match those of the object frames (Sec. \ref{subsec:constantObservation}), but the S/N was too low. Therefore, flats for the constant target dataset were constructed from flats taken at longer integration times on the following night.

In all three datasets, the time-variable background was subtracted after flat-fielding the arrays. The routine which was found to perform best was based on the ``nod subtraction'' used in the HOSTS survey pipeline, where the background frame for each data frame is made from a median of the 50 nearest frames in time to the data frame, taken during nods opposite to that of the data frame \citep{defrere2016nulling}.

Median values from a rectangular region of each flat-fielded detector channel were subtracted from the channel as a whole, due to small changes in channel bias levels that occurred on timescales on the order of a few minutes. After this step, the centers of the PSFs were determined and the IDL \textit{APER} function was used to extract aperture photometry with the use of a sky annulus that effectively made a secondary sky subtraction.

\begin{figure}
\centering
\includegraphics[width=1.0\linewidth]{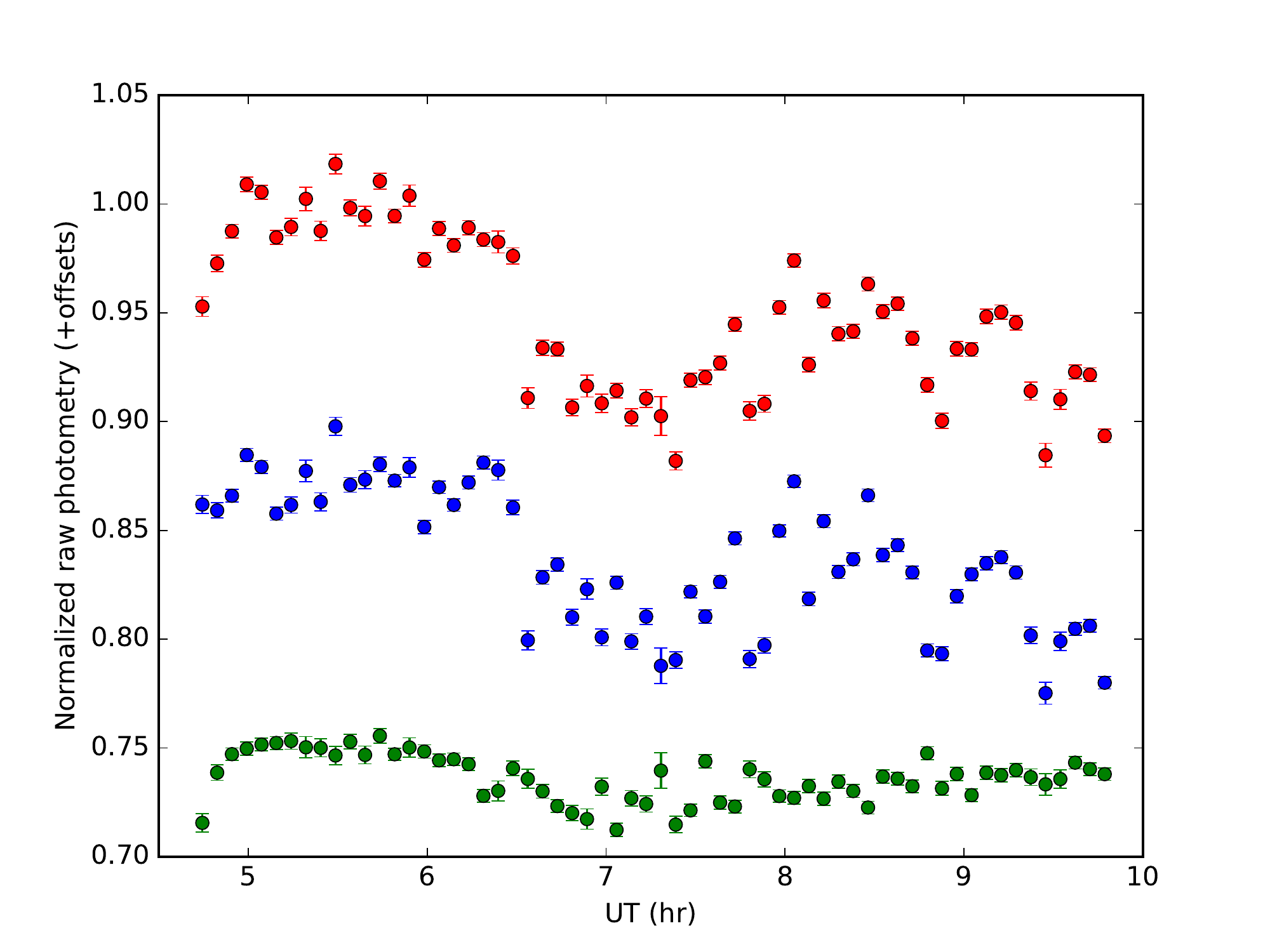}
\caption{Red: the normalized raw photometry of the science target XO-2N. Blue: the comparison star XO-2S. Green: the differential photometry found by taking the XO-2N ADU counts divided by the XO-2S ADU counts. Data points have been binned by 5 minutes and offset in y for clarity. Error bars show only $\sigma_{1}/\sqrt{N}$ without including systematic error. Note that the differential photometry reduces scatter and prevents the decrement in the raw photometry of XO-2N during the middle of the observation from masquerading as a particularly deep primary transit.} 
\label{fig:rawPhotoms}
\end{figure}

Relative photometries were generated by simply dividing the number of ADU counts from one target (the science target, if applicable) by the counts from the other. Comparison with raw photometries from the individual stars showed that the use of the comparison stars did indeed lead to substantial improvement (Fig. \ref{fig:rawPhotoms}).

\subsection{Photometry Apertures for Minimizing Noise} \label{subsec:minRedNoise}

Whereas differential photometry at shorter wavelengths calls for apertures that find the right balance between maximizing the signal and minimizing the sky footprint, the task is complicated in the thermal infrared by the increased presence of slowly-varying residual systematics, or `red' noise. Considerable effort has gone into characterizing or decorrelating its effect, particularly on exoplanet transit or secondary eclipse light curves which often require sub-percent precision (e.g., \citet{pont2006effect}, \citet{carter2009parameter}, \citet{croll2015near}). Indeed, red noise in our dataset involving the XO-2Nb primary transit led to light curves with variations of $\sim$6\% at apertures comparable to the first few Airy rings, when the expected transit depth was slightly more than 1\%.

\subsubsection{Strategy} \label{subsec:redNoiseStrategy}

`Wall-eyed' observations with the LBT are unique, and there was no pre-existing theoretical model for the red noise we should have expected to see in the photometry. We also could not choose among different combinations of comparison stars to minimize the red noise, as is possible in near-infrared observations with wider fields-of-view (e.g., \citet{narita2013multi}). Thus we were limited to varying only the photometric apertures and sky background annuli of our two targets, so as to find the combinations which struck the optimal balance between red noise variations (which dominate at small apertures) and the purely `white' photon noise (which dominates at large apertures). 

To filter out some effects of seeing bursts in the photometry, we found a running average of the seeing in a boxcar window of 2 minutes in duration, and compared this to the median of the seeing in a secondary, minute-wide trailing window. If the value in the primary window was greater than 1.5 times the standard deviation in the secondary window, the corresponding points in the photometry were flagged and removed. 

Then, adopting a strategy similar to that of \citet{croll2015near}, we mapped the amount of red noise in parameter space along dimensions of aperture sizes and sky annulus widths. Limits on the range of tested aperture radii and sky annulus widths were imposed so as to avoid having them overlap each other, and to avoid the detector edge. The inner edge of the sky annulus was also kept a fixed distance of 2 pixels beyond the aperture radius. From each image we extracted two photometric data points: one representing the number of ADU within the aperture around the science target PSF, and the other around the comparison star PSF.

For each combination of aperture size and sky annulus, we quantify the amount of residual red noise in the differential photometry. (Throughout the analysis, the differential photometry has been normalized to 1.) We make use of a quantity $\beta$ which represents the total amount of noise over that expected for pure `white' photon noise. Specifically, for a given bin size of $N$ photometric data points, we found the standard deviation $\sigma$ of the residuals of the differential photometry (see Sec. \ref{sec:results}) and $\beta$ as defined by \citet{winn2007transit} and \citet{winn2008transit} as the ratio of observed scatter over that expected for white noise, or $\beta\equiv\sigma_{obs}/\sigma_{exp}$. For $M$ bins of $N$ data points, we use Eqn. 2 of \citet{winn2008transit},

\begin{equation}\label{eq:sigma_N}
\sigma_{exp}=\frac{\sigma_{1}}{\sqrt{N}}\sqrt{\frac{M}{M-1}}
\end{equation}

\noindent
where $\sigma_{exp}$ is the expected standard deviation of Gaussian-distributed photometric data points after binning, and $\sigma_{1}$ is that of the unbinned photometry. The $\sqrt{M/(M-1)}$ is Bessel's correction for obtaining the unbiased standard deviation. It should be noted that Eqn. \ref{eq:sigma_N} is referring to the scatter in photometric data points after having extracted each data point from each image. The detector readouts themselves are not being stacked or binned.

In cases where the number of data points cannot be evenly divided by bins of a given size, we use fractional numbers of bins in Eqn. \ref{eq:sigma_N}, and find the standard deviations while sliding the bins across the dataset. We then average the standard deviations to get a single value corresponding to a single bin size.

\subsubsection{Red Noise in Aperture-annulus Space} \label{subsec:redNoiseBehavior}

\begin{figure}
\centering
\includegraphics[width=1.0\linewidth,trim={2.2cm 7cm 2.5cm 7.7cm},clip=true]{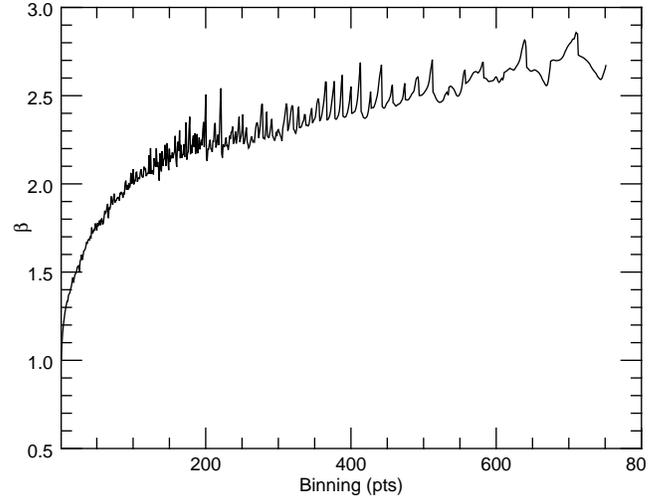}
\caption{The behavior of $\beta$ in the primary transit dataset as a function of the number of photometric data points in each bin, with an aperture radius of 42 pixels and an annulus width of 36 pixels.} 
\label{fig:beta}
\end{figure}

The value of $\beta$ usually tended to converge to a linear (or constant) trend, aside from a shark-tooth aliasing pattern with a period corresponding to bin sizes that can be divided, or nearly divided, into the number of data points (Fig. \ref{fig:beta}). In all three datasets, a rough convergence of $\beta$ to a constant or linear trend appears to occur by the time the bin size is 200 points. 

We calculate an average $\bar{\beta}$ from $\beta$ values corresponding to bin sizes between 200 and 400 points in the constant target data, and between 200 and 751 points in the primary transit and secondary eclipse data. The upper limit of 751 data points represents $\sim$15 minutes of observation in the primary transit data, beyond which $\beta$ tends to become racked with aliasing artifacts. The smaller range of bin sizes in the constant target data was used because $\beta$ appeared to converge at these sizes, then to precipitously descend beyond a bin size of 400 points. This range in bin size then provides an upper limit for $\beta$. In order to determine the lowest obtainable scatter in the datasets, we adopt the standard deviation of the photometry after it has been binned at the largest bin size used to find $\beta$, and then multiply the standard deviation by $\beta$ to include the effects of red noise. 

\begin{figure*}
\centering
\includegraphics[width=15cm, trim={0cm 2.5cm 0cm 3cm},clip=true]{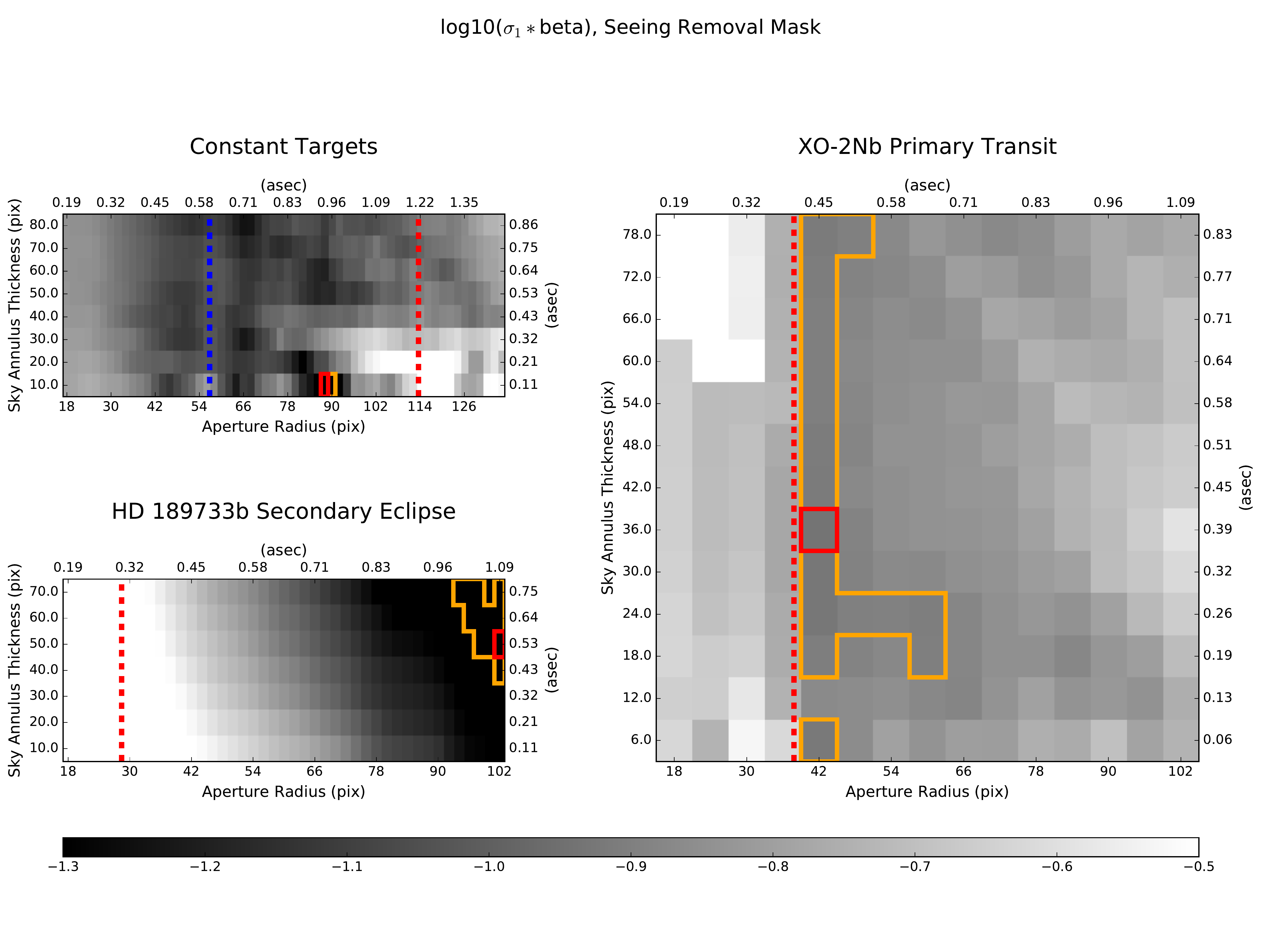} 
\caption{Plots of $log_{10}(\sigma\bar{\beta})$ for the three datasets. Red-bordered boxes indicate the lowest values in the plot, and the orange border (sometimes overlapped by the red border) indicates regions that are within 10\% of the lowest value. Dashed lines represent approximate locations of control radii as described in Sec. \ref{subsec:minRedNoise}, with a $\pm6$ \% spread due to the wavelength range of the $L_{S}$ filter. Blue dashed lines are for the control radius of the left aperture, and red is for the right aperture. In the primary transit dataset, the two overlap. In the secondary eclipse dataset, the left-aperture control radius is outside the plot at 1.2 asec. The color scale in all plots is the same. To bring out similar levels of contrast in each plot, the scale has been truncated (i.e.,  the minimum (black) and maximum (white) ends of the scale should be read as $\leq-1.3$ and $\geq-0.5$.)} 
\label{fig:sigmaBetaSpace}
\end{figure*}

Fig. \ref{fig:sigmaBetaSpace} shows the aperture-annulus space mapped. Due to its size, the primary transit dataset was mapped at coarser resolution than that of the other two datasets. For each dataset we selected the combination of aperture and annulus sizes which produced the lowest values of $\sigma_{1}\bar{\beta}$. In Fig. \ref{fig:sigmaBetaSpace} we also use dashed red and blue lines to indicate the control radii, or the radii extending from the PSFs within which the AO correction is active. This is a finite radius which is imposed by virtue of the fact that wavefront aberrations can only be sampled down to the Nyquist limit for a given number of subapertures across the pupil plane (e.g., \citet{poyneer2004spatially}). Beyond the control radius, photons from higher-spatial-frequency modes appear as a seeing-limited halo. 

\subsection{Systematic Error Contributions}
\label{subsec:systematics}

Dividing the science target ADU counts by the comparison star ADU counts led to improvements in the photometries. However, strong instrumental or environmental systematic errors persisted. We proceeded to investigate various possible underlying causes, with a view to identifying and removing the effects.

\subsubsection{Background}

These observations are background-limited, so we proceeded to investigate the time dependency and dominant sources of background flux. A Sky Quality Meter by Unihedron measured the $V$-band sky brightness during the primary transit and secondary eclipse observations. The median values were 19.4 and 21.4 mag asec$^{-2}$, respectively, though the background in $L$ is expected to be $\sim$15 mag asec$^{-2}$ brighter \citep{longair2011high}. Using the primary transit dataset, the median stellar ADU pix$^{-2}$ sec$^{-1}$ within the photometric apertures eventually chosen (see Sec. \ref{subsec:minRedNoise}), and $K-L=0.06$ for late-G-type stars \citep{tokunaga2000allen}, we adopt a conversion of $\sim$2222 ADU sec$^{-1}$ mJy$^{-1}$. 

Assuming the response of the detector is the same in all datasets, we find median $L_{S}$-band background levels (which consist of overlapping sky backgrounds from both apertures) of 20, 9, and 2 Jy asec$^{-2}$ in the constant target, primary transit, and secondary eclipse observations, respectively. The target stars had brightnesses of $<$1 Jy in $L_{S}$. Thus the two overlapped and time-variable $L_{S}$-band sky backgrounds represented a pedestal of the same order brightness of the PSFs. 

The time-dependency of the background is probably due to a component of the background which does not originate from the telescope or dome. Chamber temperature variations were always measured to be $<3^{\circ}$ C, and the corresponding change of the ambient emission at $L_{S}$ would have been $<15$ \%. In fact, the background levels and calculated $L_{S}$-band dome emissivity are anticorrelated in the constant target and primary transit datasets.

\begin{figure*}
\begin{minipage}{0.3\textwidth}
	\centering
	Constant targets
\end{minipage}
\hspace{0.15in}
\begin{minipage}{0.3\textwidth}
	\centering
	Primary transit
\end{minipage}
\hspace{0.15in}
\begin{minipage}{0.3\textwidth}
	\centering
	Secondary eclipse
\end{minipage}
\centering
\includegraphics[height=5cm,trim={0.5cm 0cm 2cm 1cm},clip=true]{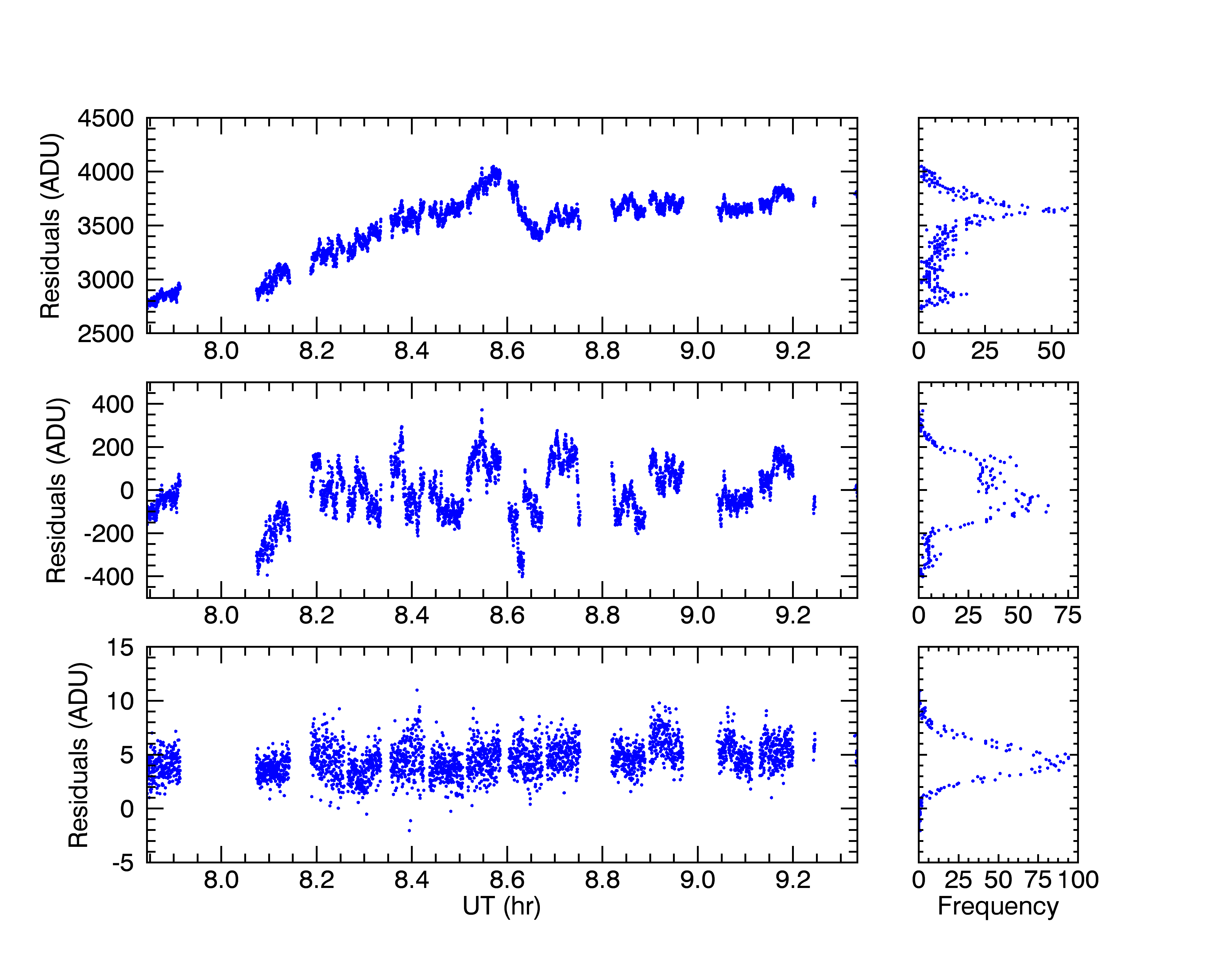}
\hspace{0.2cm}
\includegraphics[height=5cm,trim={0.5cm 0cm 2cm 1cm},clip=true]{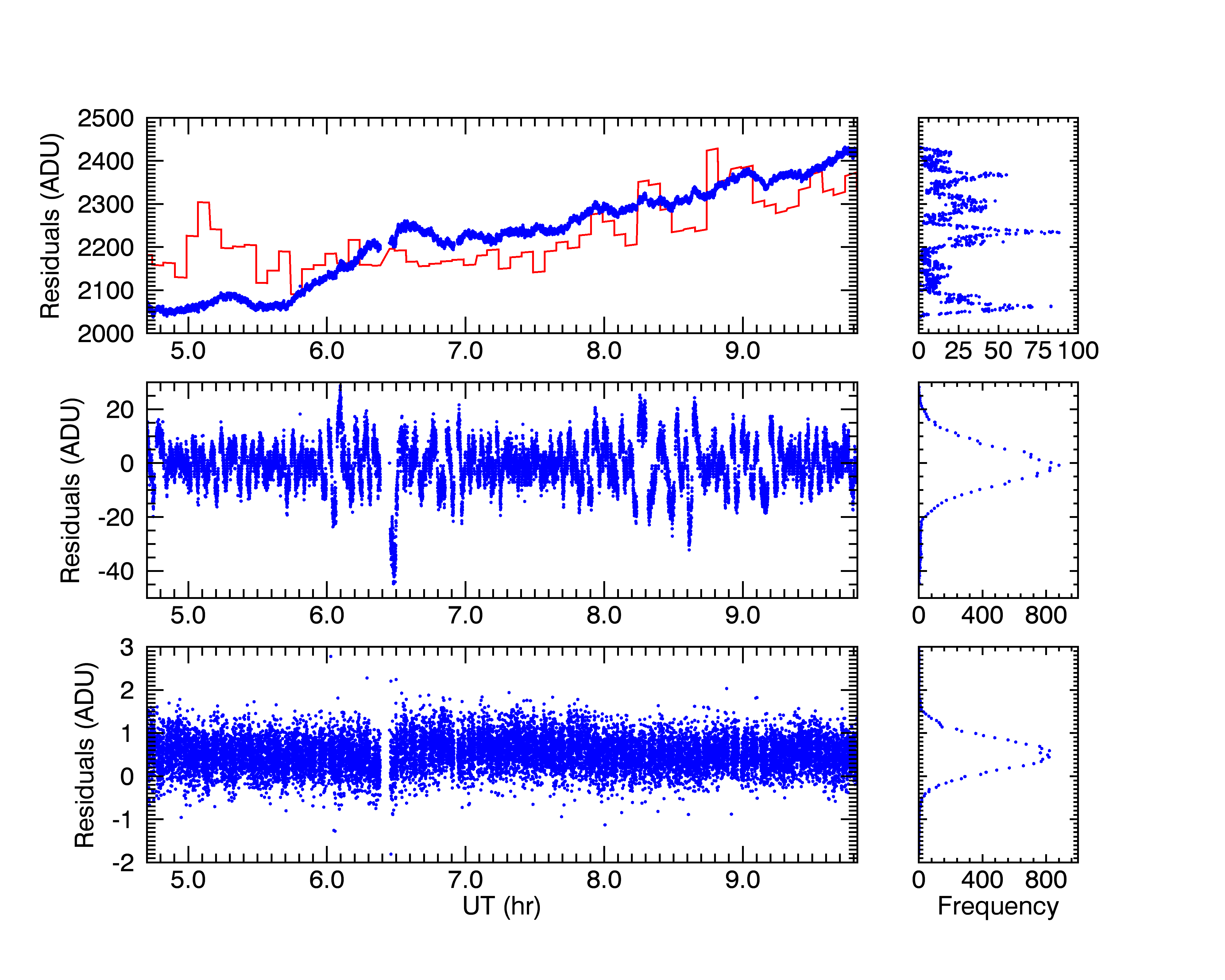}
\hspace{0.2cm}
\includegraphics[height=5cm,trim={0.5cm 0cm 2cm 1cm},clip=true]{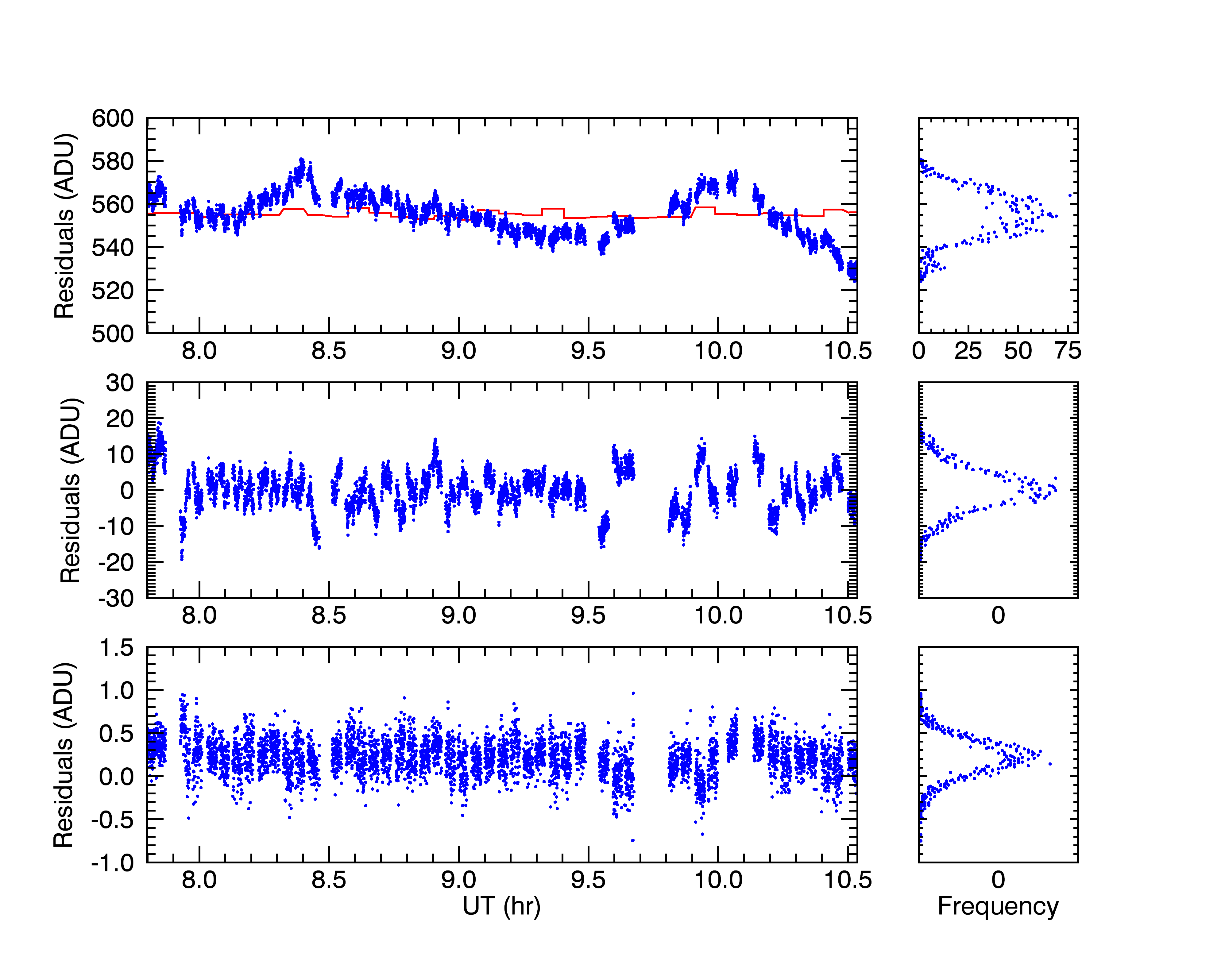}
\caption{The background levels at different stages of the reduction, with rotated histograms. Left to right blocks: the observations of the constant targets, the primary transit, and the secondary eclipse. Top to bottom rows: the original background level, the background level after nod subtraction, and the final background level as determined by the aperture background annuli around the PSF. (Background values for the latter are averages between the two PSFs.) In two of the datasets, the red lines in the top panels show the model sky emissivity based on the changing airmass and precipitable water vapor as modeled by Eqn. \ref{eqn:toyModel}.}
\label{fig:backgroundLevels}
\end{figure*}

We take this to mean that the large-amplitude, time-dependent variations in the background with a timescale on the order of minutes or hours (Fig. \ref{fig:backgroundLevels}) are from a component of the background due to the atmosphere. This allows us to fit a linear model

\begin{equation}
B\{E\} = c_{tel} + c_{atm}E\{a,PWV;t\}
\label{eqn:toyModel}
\end{equation}

\noindent 
where $B$ is a model background level and $E\{a,PWV;t\}$ is a model sky emissivity as an explicit function of airmass $a$ and precipitable water vapor $PWV$, and an implicit function of time $t$. The $c_{tel}$ represents the time-independent contribution from the telescope, which includes thermal contributions from dome-temperature sources such as the primary, secondary, and tertiary mirror surfaces, dust on the primary mirror, and the secondary and tertiary mirror swing arms. The constant $c_{atm}$ involves the effect of the atmosphere. We find $E\{a,PWV;t\}$ by making interpolations of emissivity spectra based on changing airmass and precipitable water vapor levels. The grid of spectra is provided by Gemini Observatory,\footnote{\url{http://www.gemini.edu/sciops/telescopes-and-sites/observing-condition-constraints/ir-background-spectra}} based in part on model transmissivity data by \citet{lord1992nasa} (e.g., Fig. \ref{fig:maunakeaemission}). 

The best fits to the empirical backgrounds of the primary transit and secondary eclipse datasets are shown as the red lines in Fig. \ref{fig:backgroundLevels}. In the case of the primary transit observation, comparison of $c_{tel}$ to the total background levels $B$ in Eqn. \ref{eqn:toyModel} suggests that $\sim$50-60 \% of the background emission is due to the telescope, and most of the rest is due to the atmosphere (Table \ref{table:backgroundContrib}). The fit to the background of the secondary eclipse data in Fig. \ref{fig:backgroundLevels} shows less evidence for correspondence between sky background levels and precipitable water vapor. However, that observation was over a shorter time baseline and a smaller change in airmass, both of which lead to poorer sampling of possible correlations between the empirical background levels and the model in Eqn. \ref{eqn:toyModel}. 

\begin{table}
\begin{center}
\caption{Transit observation background contributors} 
\label{table:backgroundContrib}
\begin{tabular}{ l c c c }
\hline
\textit{Source}			& 	\textit{Timescale}		&		\textit{Effect}	 \\
 \hline	
Telescope		& 		$\sim$constant	& 	$\gtrsim$ half 	 \\	
\hline
Sky		& 		$\sim$hrs 		&  	$\lesssim$ half 	 \\	
\hline
Water vapor variations		& 		$\lesssim$1/2 hr 		&  	$\lesssim$ few percent 	 \\	
\end{tabular}
\noindent
\end{center}
\end{table}

\begin{table}
\begin{center}
\caption{Transit observation noise contributors} 
\label{table:noiseContribs}
\begin{tabular}{ l c c c }
\hline
\textit{Source}			& 	\textit{Timescale}		&		\textit{Effect}	 \\
 \hline	
Wind-borne (?)		& 		$\sim$secs to $\sim$hrs	& 	dominant 	 \\	
\hline
Seeing rises		& 		$\sim$secs to $\sim$mins 		&  	sporadic 	 \\	
\hline
Detector bias variations		& 		$\sim$mins 		&  	negligible 	 \\	
\end{tabular}
\noindent
\end{center}
\end{table}

\subsubsection{PSF Movement}

It was noted that the PSFs sometimes migrated across several pixels on the detector array, especially in the primary transit and secondary eclipse datasets. The rate at which the PSFs moved sometimes correlated with elevation, which may suggest flexure downstream of the telescope. Other contributing factors may have included the accumulation of nod position errors over time, and the rotation of the nod vector. However, comparison of the photometry before and after linearization indicated that differential nonlinearity effects among the pixels had a negligible effect on the photometry, whose systematics were evidently dominated by other sources.

\subsubsection{Wind-correlated Effective Transmission: $\Delta Q$}
\label{subsubsec:deltaQ}

We considered the possibility that the lines-of-sight of the two apertures were sampling different regions of a time-varying gradient of some unknown quantity $Q$, which can be thought of as an effective transmissivity. This gradient swivels around over the telescope as a function of wind direction, and the difference of the quantity over the two apertures is proportional to the dot product of the wind direction vector $\hat{v_{w}}(t)$ and the vector $\hat{B}(t)$ along the baseline from the center of the left aperture to the center of the right aperture. Dropping the $t$ for notational simplicity, $\Delta Q \propto \hat{v_{w}}\cdot \hat{B}$. In the limit that the wind $\overrightarrow{v_{w}}$ moves over the apertures normal to $\hat{B}$, $\hat{v_{w}}\cdot \hat{B}$ goes to zero because the gradient over the two apertures is the same. If the wind is blowing directly along the aperture baseline, then line-of-sight variations will be strongest. It should be noted that the wind is measured from the roof, and does not contain information about the full wind structure along the lines-of-sight. Our use of the rooftop wind is meant only to serve as an approximate substitute for a more complete wind structure.

\subsection{Parametric Model} \label{subsec:fitting}

We fit the relation $F_{tot}=F_{sys}F_{mod}$ to each of our three sets of differential photometry, where $F_{sys}$ represents the wind-borne contribution to the systematics. For the constant target and secondary eclipse datasets, the model flux $F_{mod}$ is simply set to 1. For the XO-2Nb primary transit, we applied a transit model $T$ based on the equations in \citet{mandel2002analytic}, using limb-darkening coefficients for a quadratic limb-darkening law in Bessell-$L$ band corresponding to a star of $T_{eff}=5250$ K and $log(g)=4.5$ \citep{howarth2011new}. The model flux can be written as

\begin{equation}
F_{mod}(t) = T\left(t,t_{TOC},T_{tot},R_{P}/R_{S},b\right)
\label{eqn:transitModel}
\end{equation}

\noindent
where $t$ is time, $t_{TOC}$ is the transit time-of-center, $T_{tot}$ is the time between first and last times of contact between the disks of the planet and the host star, $R_{P}/R_{S}$ is the ratio of the planet to stellar radii, and $b=0.16$ \citep{torres2008improved} is the minimum impact parameter.

We model the systematics as 

\begin{equation}
F_{sys}(t) = \left[1 + c_{1} + \left(\frac{c_{2}t}{hr}\right)\right](1 + c_{3} + c_{4}\Phi\{\overrightarrow{v_{w}},\hat{B};t\})
\label{eqn:systemEqn}
\end{equation}

\noindent
where $c_{1}$, $c_{2}$, $c_{3}$, and $c_{4}$ are constants. The term in square brackets is the simplest explicitly time-dependent systematics term: a linear function of $t$. Due to the short time baseline of the constant target and secondary eclipse observations, we set $c_{1}\equiv c_{2}\equiv0$ for those two datasets to avoid overfitting.

Correlations between the photometry of the primary transit dataset and the pattern of wind flow across the two apertures (Sec. \ref{subsubsec:deltaQ}) motivated the term in curved brackets in Eqn. \ref{eqn:systemEqn}, where $\Phi$ is meant to encapsulate the changing effect of wind flow. The term in curved brackets is linear in $\Phi$ for simplicity, and the additive offset $c_{3}$ allows for the effect of wind which is not captured in the time-dependent variation.

We allow three possibilities for $\Phi$ by letting it be either proportional to the dot products $\hat{v_{w}}\cdot \hat{B}$, $\overrightarrow{v_{w}}\cdot \hat{B}$, or zero. The caret symbols indicate unit vectors which are normalized to a length of 1. In the case of $\Phi=0$ we set $c_{3}\equiv 0$. In what follows, the amplitude of the quantity $\hat{v_{w}}\cdot \hat{B}$ has been normalized such that the maximum and minimum values of a smoothed plot of this quantity differ by 1, so as to avoid numerical problems during the MCMC. No modification is made for telescope elevation changes.

\subsection{Parametric MCMC Analysis} \label{subsec:mcmc}

The posterior distribution of the probability in a parameter space can be expressed using Bayes' theorem

\begin{equation}
P(\textbf{p}|\textbf{y})\propto P(\textbf{p})P(\textbf{y}|\textbf{p})
\end{equation}

\noindent
for model transit parameters \textbf{p} and data points \textbf{y}. We sampled this posterior using the open-source \texttt{emcee} Python MCMC module \citep{foreman2013emcee}, where Gaussian offsets to an initial Levenberg-Marquardt fit were used as the starting points for 50 chains of $10^{5}$ post-burn-in links. Error bars on the photometric data points were set to $\sigma_{1}\bar{\beta}$. 

\begin{table}
\begin{flushleft}
\caption{MCMC Priors$^{a}$}
\label{table:priors}
\begin{tabular}{ p{.2\linewidth} p{.19\linewidth} p{.23\linewidth} p{.2\linewidth} }
\hline
\textit{Quantity}          &       \textit{Prior type}            &               \textit{Condition}    &      \textit{Units}     \\
 \hline
\textbf{$t_{TOC}$}          &               Gaussian           &       $7.5 \pm0.5$           &       hr (UT)                                           \\
\hline
\textbf{$T_{tot}$}          &               Gaussian            &       $2.67 \pm1.00$            &         hr                                         \\
\hline
\textbf{$R_{P}/R_{S}$}          &               Gaussian            &       $0.103 \pm0.010$            &                                                   \\
\hline
\textbf{$c_{1}$}          &               Flat            &       $\lvert c_{1}\rvert<1$            &                                                  \\
\hline
\textbf{$c_{2}$}          &               Flat            &       $\lvert c_{2}\rvert<1$            &                                                  \\
\hline
\textbf{$c_{3}$}            &               Flat              &       $\lvert c_{3}\rvert<1$          &           \\
\hline
\textbf{$c_{4}$}            &               Flat              &       $\lvert c_{4}\rvert<1$          &           \\
\end{tabular}
\noindent
\tablenotetext{a}{Coefficients $c_{i}$ are relevant to all three datasets, but planetary transit parameters are specific to the XO-2Nb primary transit dataset only.}
\end{flushleft}
\end{table}

\begin{figure*}
\includegraphics[width=17.5cm]{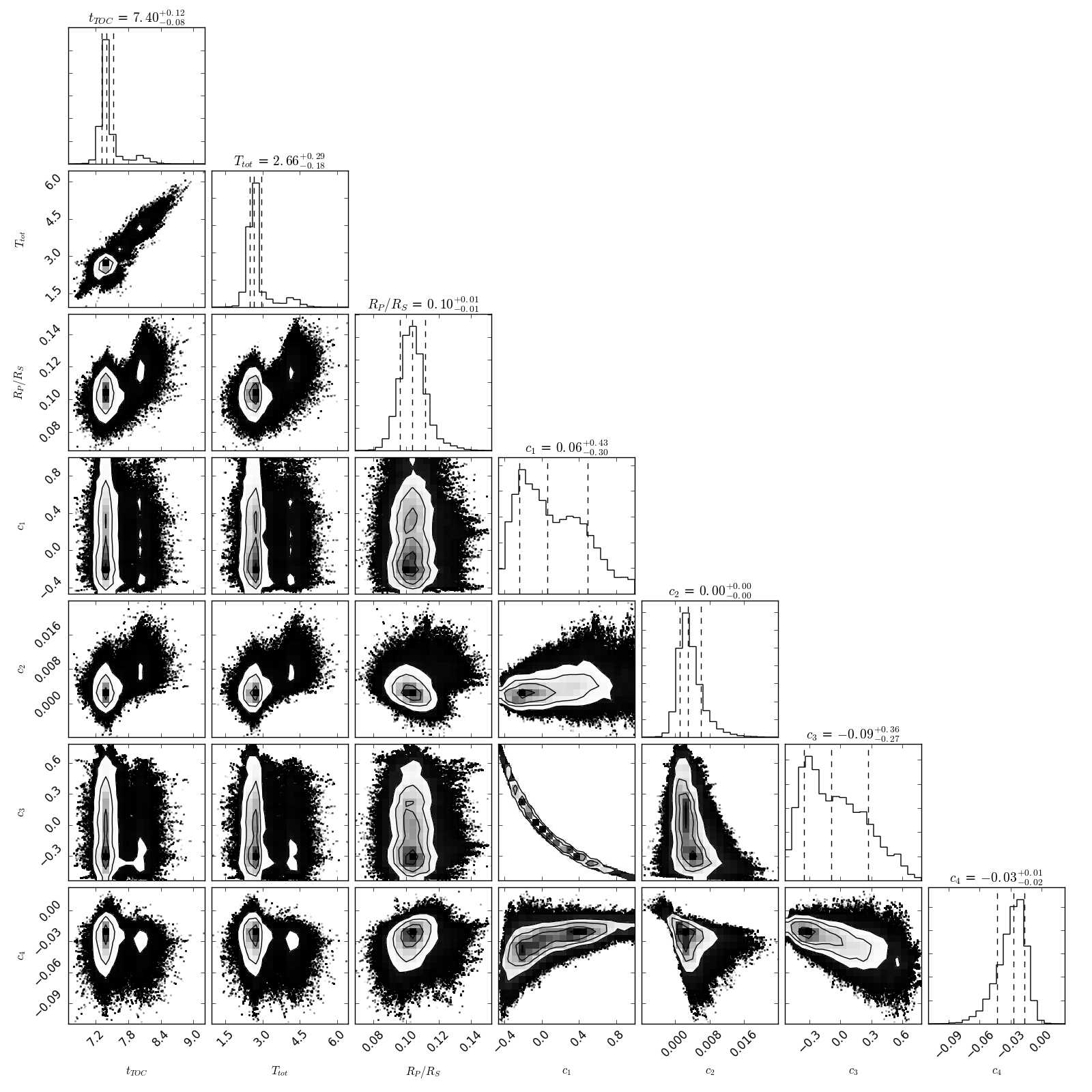}
\caption{Primary transit parameters, with systematics term $\Phi \propto \hat{v_{w}}\cdot \hat{B}$. Vertical dashed lines in the 1D histograms represent the median and $\pm1\sigma$ boundaries. Contours on the 2D histograms correspond to 2D 1$\sigma$, 1.5$\sigma$, and 2$\sigma$ levels.} 
\label{fig:feb2015corner}
\end{figure*}

The freely-varying parameters included the background coefficients, and, in the primary transit dataset, $t_{TOC}$, $T_{tot}$, and $R_{P}/R_{S}$. (The initial Levenberg-Marquardt fit used a fixed transit depth.) We used uninformative flat `top hat' priors for the background coefficients, and Gaussians for transit parameters with means and standard deviations as indicated in Table \ref{table:priors}. Results are shown in Figs. \ref{fig:feb2015corner} to \ref{fig:june2015MCMCsamples}, and are discussed in detail in Sec. \ref{sec:results}.

\newpage

\section{Results} \label{sec:results}

Coefficients for the parameters of the best-fitting models involving $\Phi=0$, $\Phi \propto \hat{v_{w}}\cdot \hat{B}$, and $\Phi\propto\overrightarrow{v_{w}}\cdot \hat{B}$ were taken to be the median values from the posterior samples. In the case of the XO-2Nb primary transit, the median coefficients and parameters from Eqns. \ref{eqn:transitModel} and \ref{eqn:systemEqn} are tabulated in Table \ref{tab:transitResults}, where stated errors correspond to the equivalent of 1$\sigma$ errors.

\floattable
\begin{deluxetable}{c|l|ccc|cccc}
\rotate
\tablewidth{200pt}
\tabletypesize{\scriptsize}
\tablecaption{Median transit and systematics parameters with 1$\sigma$ errors$^{a}$\label{tab:transitResults}}
\tablehead{
\colhead{Dataset} & 
\colhead{Fit or reference} & 
\colhead{$t_{TOC}- 2457061$ JD} & 
\colhead{$T_{tot}$ (hr)}& 
\colhead{$R_{P}/R_{S}$} &
\colhead{$c_{1}$} & 
\colhead{$c_{2}$} & 
\colhead{$c_{3}$} & 
\colhead{$c_{4}$} 
}
\startdata
\multirow{3}{*}{Constant targets}
& MCMC, $\Phi=0$ & $-$ & $-$ & $-$ & $-0.00075^{+0.00089}_{-0.00089}$ & $-$ & $-$ & $-$ \\
& MCMC, $\Phi\propto\hat{v_{w}}\cdot \hat{B}$ & $-$ & $-$ & $-$ & $-0.14^{+0.48}_{-0.34}$ & $-$ & $0.16^{+0.75}_{-0.42}$ & $0.0022^{+0.0026}_{-0.0016}$ \\
& MCMC, $\Phi\propto\overrightarrow{v_{w}}\cdot \hat{B}$ & $-$ & $-$ & $-$ & $-0.31^{+0.22}_{-0.26}$ & $-$ & $0.44^{+0.89}_{-0.52}$ & $6^{+34}_{-32}$ ($\times 10^{-5}$) \\
\hline
\multirow{6}{*}{XO-2Nb transit}
& LM, $\Phi=0$ 		& 0.8164 		& 2.6784 (fixed) 		& 0.1030 (fixed) & 0.00504143 & -0.00166390 & $-$ & $-$ \\ 
& MCMC, $\Phi=0$ 	& $0.8090^{+0.0030}_{-0.0028}$ 	& $2.61^{+0.14}_{-0.17}$ 	& $0.1132^{+0.0062}_{-0.0065}$ & $0.22^{+0.49}_{-0.41}$ & $-0.0024^{+0.0009}_{-0.0013}$ & $-0.17^{+0.41}_{-0.24}$ & $-$ \\
& MCMC, $\Phi\propto\hat{v_{w}}\cdot \hat{B}$ 	& $0.8081^{+0.0052}_{-0.0034}$ 	& $2.66^{+0.29}_{-0.18}$	& $0.1036^{+0.0080}_{-0.0072}$ & $0.06^{+0.43}_{-0.30}$ & $0.0031^{+0.0030}_{-0.0020}$ & $-0.09^{+0.36}_{-0.27}$ & $-0.027^{+0.011}_{-0.016}$ \\
& MCMC, $\Phi\propto\overrightarrow{v_{w}}\cdot \hat{B}$ 	& $0.8094^{+0.0204}_{-0.0034}$ 	& $2.66^{+1.11}_{-0.22}$ 	& $0.1034^{+0.0092}_{-0.0081}$ & $0.11^{+0.40}_{-0.31}$ & $0.0027^{+0.0035}_{-0.0022}$ & $-0.13^{+0.33}_{-0.23}$ & $-0.0029^{+0.0012}_{-0.0016}$ \\
& \citet{zellem2015xo}, Bessel-$U$ & $0.8169^{+0.0023}_{-0.0040}$  & (fixed) & $0.1087^{+0.0033}_{-0.0051}$ & $-$ & $-$ & $-$ & $-$ \\
& \citet{zellem2015xo}, Harris-$B$  & $0.8164^{+0.0025}_{-0.0007}$  & (fixed) & $0.1030^{+0.0028}_{-0.0012}$ & $-$ & $-$ & $-$ & $-$ \\
\hline
\multirow{3}{*}{HD 189733b sec. eclipse}
& MCMC, $\Phi=0$ & $-$ & $-$ & $-$ & $0.000255^{+0.00061}_{-0.00061}$ & $-$ & $-$ & $-$ \\
& MCMC, $\Phi\propto\hat{v_{w}}\cdot \hat{B}$ & $-$ & $-$ & $-$ & $-0.18^{+0.43}_{-0.31}$ & $-$ & $0.22^{+0.75}_{-0.42}$ & $-0.0151^{+0.0056}_{-0.0099}$ \\ 
& MCMC, $\Phi\propto\overrightarrow{v_{w}}\cdot \hat{B}$ & $-$ & $-$ & $-$ & $-0.12^{+0.45}_{-0.27}$ & $-$ & $0.14^{+0.49}_{-0.38}$ & $-0.00266^{+0.00095}_{-0.00142}$ \\
\enddata
\tablenotetext{a}{Transit and noise model parameters are defined in Sec. \ref{subsec:fitting}.}
\end{deluxetable}

\floattable
\begin{deluxetable}{c|l|ccccccccc}
\rotate
\tablewidth{200pt}
\tabletypesize{\scriptsize}
\tablecaption{Photometric Red Noise\label{table:redNoiseResults}}
\tablehead{
\colhead{Dataset} & 
\colhead{Fit} & 
\colhead{Aperture (pix, '', $\lambda_{c}/D$)$^{a}$} & 
\colhead{Annulus (pix, '', $\lambda_{c}/D$)$^{a}$} & 
\colhead{$\sigma_{1}$$^{b}$} & 
\colhead{$\bar{\beta}$$^{c}$} & 
\colhead{$\sigma_{1}'\equiv\sigma_{1}\bar{\beta}$$^{d}$} & 
\colhead{$N_{max}$$^{e}$} & 
\colhead{$N_{data}/N_{max}$$^{f}$} & 
\colhead{$\sigma_{Nmax}'$ ($10^{-3}$)$^{g}$} & 
\colhead{$\sigma_{Nmax}'$ (mmag)}
}
\startdata
\multirow{4}{*}{Constant targets}
& LM, $\Phi=0$ & 88, 0.94, 11.6 & 10, 0.11, 1.3 & 0.030 & 1.214 & 0.036 & 400 & 4.7 & 2.1 & 2.2\\
& MCMC, $\Phi=0$ & 88, 0.94, 11.6 & 10, 0.11, 1.3 & 0.030 & 1.214 & 0.036 & 400 & 4.7 & 2.1 & 2.2\\
& MCMC, $\Phi=\hat{v_{w}}\cdot \hat{B}$ & 88, 0.94, 11.6 & 10, 0.11, 1.3 & 0.030 & 1.178 & 0.035 & 400 & 4.7 & 2.0 & 2.2\\
& MCMC, $\Phi=\overrightarrow{v_{w}}\cdot \hat{B}$ & 88, 0.94, 11.6 & 10, 0.11, 1.3 & 0.030 & 1.206 & 0.036 & 400 & 4.7 & 2.0 & 2.2\\
\hline
\multirow{4}{*}{XO-2Nb transit}
& LM, $\Phi=0$ & 42, 0.45, 5.5 & 36, 0.38, 4.7 & 0.047 & 2.496 & 0.116 & 751 & 17.1 & 4.4 & 4.8\\ 
& MCMC, $\Phi=0$ & 42, 0.45, 5.5 & 36, 0.38, 4.7 & 0.046 & 1.893 & 0.088 & 751 & 17.1 & 3.3 & 3.6\\
& MCMC, $\Phi=\hat{v_{w}}\cdot \hat{B}$ & 42, 0.45, 5.5 & 36, 0.38, 4.7 & 0.046 & 1.850 & 0.086 & 751 & 17.1 & 3.2 & 3.5\\
& MCMC, $\Phi=\overrightarrow{v_{w}}\cdot \hat{B}$ & 42, 0.45, 5.5 & 36, 0.38, 4.7 & 0.046 & 1.893 & 0.088 & 751 & 17.1 & 3.3 & 3.6\\
\hline
\multirow{4}{*}{HD 189733b sec. eclipse}
& LM, $\Phi=0$ & 102, 1.09, 13.4 & 50, 0.53, 6.6 & 0.015 & 2.143 & 0.032 & 751 & 4.1 & 1.3 & 1.5\\ 
& MCMC, $\Phi=0$ & 102, 1.09, 13.4 & 50, 0.53, 6.6 & 0.015 & 2.143 & 0.032 & 751 & 4.1 & 1.3 & 1.5\\
& MCMC, $\Phi=\hat{v_{w}}\cdot \hat{B}$ & 102, 1.09, 13.4 & 50, 0.53, 6.6 & 0.015 & 1.968 & 0.029 & 751 & 4.1 & 1.2 & 1.3\\ 
& MCMC, $\Phi=\overrightarrow{v_{w}}\cdot \hat{B}$ & 102, 1.09, 13.4 & 50, 0.53, 6.6 & 0.015 & 1.880 & 0.028 & 751 & 4.1 & 1.2 & 1.3\\
\enddata
\tablenotetext{a}{These are listed in three equivalent units: pixels, asec, and $\lambda_{c}/D$ (where $\lambda_{c}/D=3.3\mu$m/8.4m).}
\tablenotetext{b}{The standard deviation of the unbinned photometry (or equivalently, binning with $N=1$ data points in each bin). (See Sec. \ref{subsec:lbti}.)}
\tablenotetext{c}{The average ratio of the observed scatter to that expected for purely white noise. (See Sec. \ref{subsec:redNoiseBehavior}.)}
\tablenotetext{d}{The scatter of the unbinned photometry, rescaled to include red noise. (See Sec. \ref{subsec:redNoiseBehavior}.)}
\tablenotetext{e}{The maximum number of data points in the bins used to calculate $\bar{\beta}$. (See Sec. \ref{subsec:redNoiseBehavior}.)}
\tablenotetext{f}{The ratio of total unmasked data points in the dataset to the maximum number of data points in a single bin.}
\tablenotetext{g}{The photometric scatter after binning the data points at the maximum bin size, and including scaling to account for red noise.}
\end{deluxetable}

\begin{figure*}
\begin{centering}
\begin{minipage}{0.333\textwidth}
	\centering
	$\Phi=0$
\end{minipage}
\begin{minipage}{0.333\textwidth}
	\centering
	$\Phi \propto \hat{v_{w}}\cdot \hat{B}$ 
\end{minipage}
\begin{minipage}{0.333\textwidth}
	\centering
	\hspace{-0.1in}
	$\Phi \propto \overrightarrow{v_{w}}\cdot \hat{B}$
\end{minipage}
\vspace{-0.1in}
\end{centering}\\
\includegraphics[width=6cm]{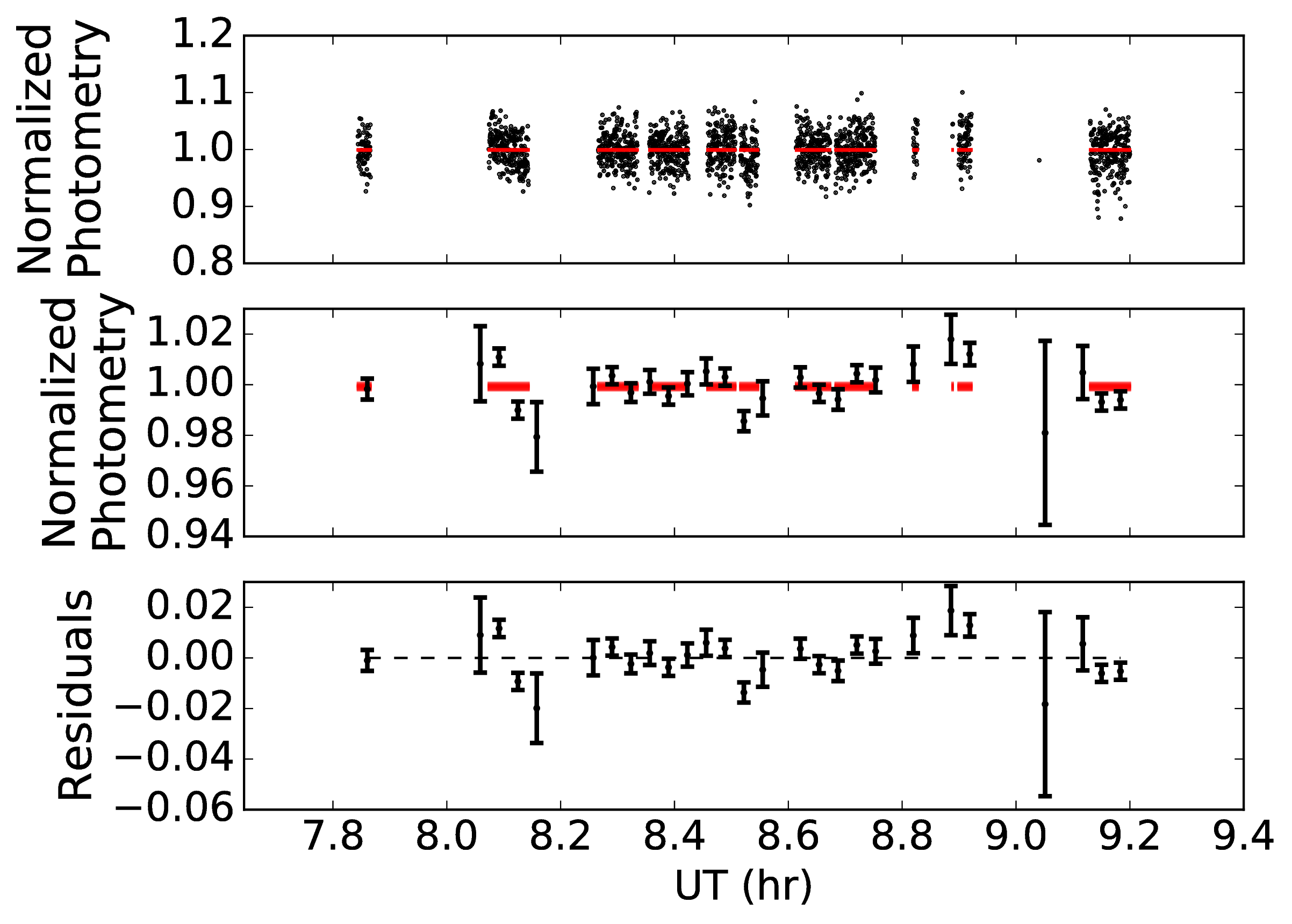}
\includegraphics[width=6cm]{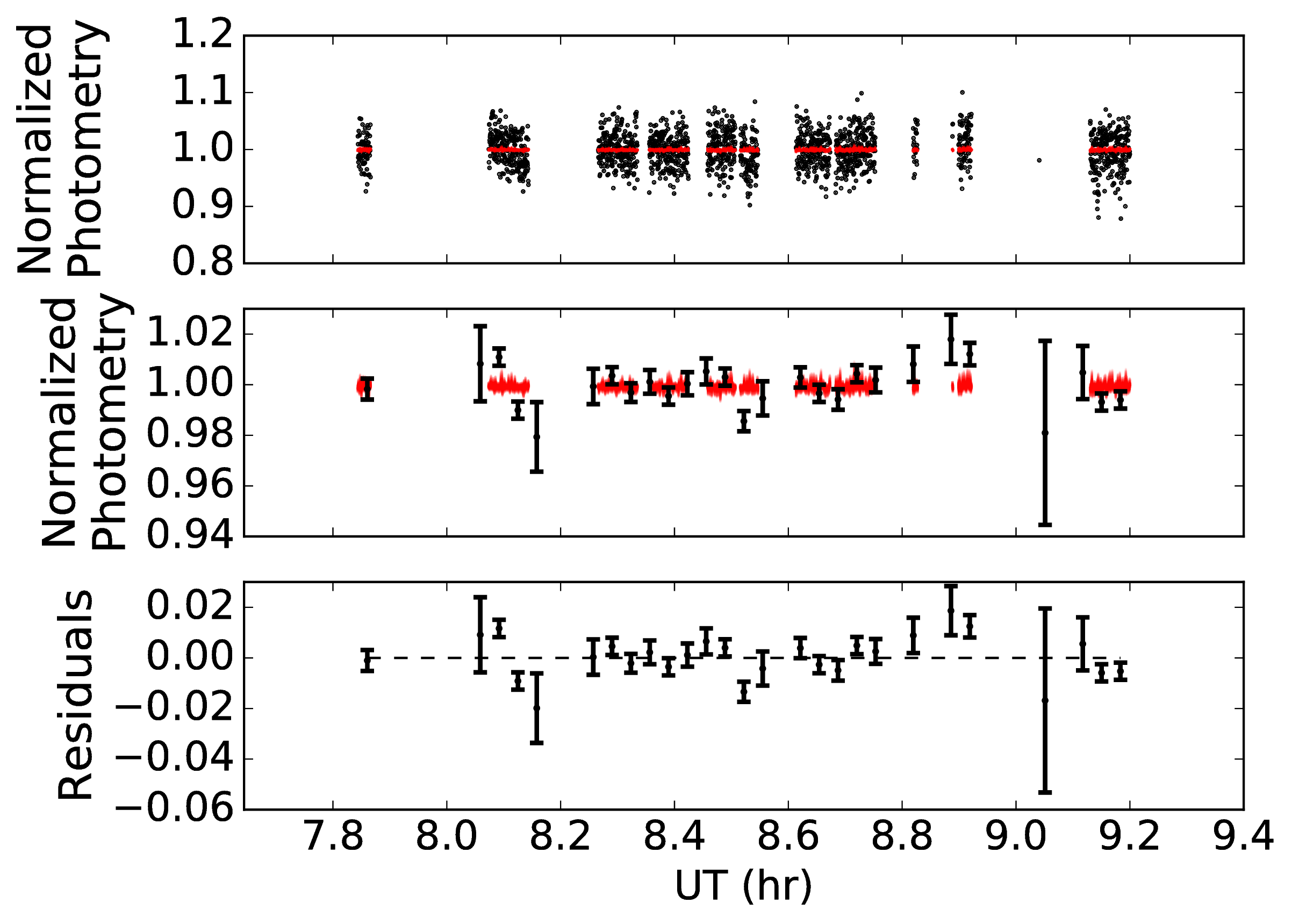}
\includegraphics[width=6cm]{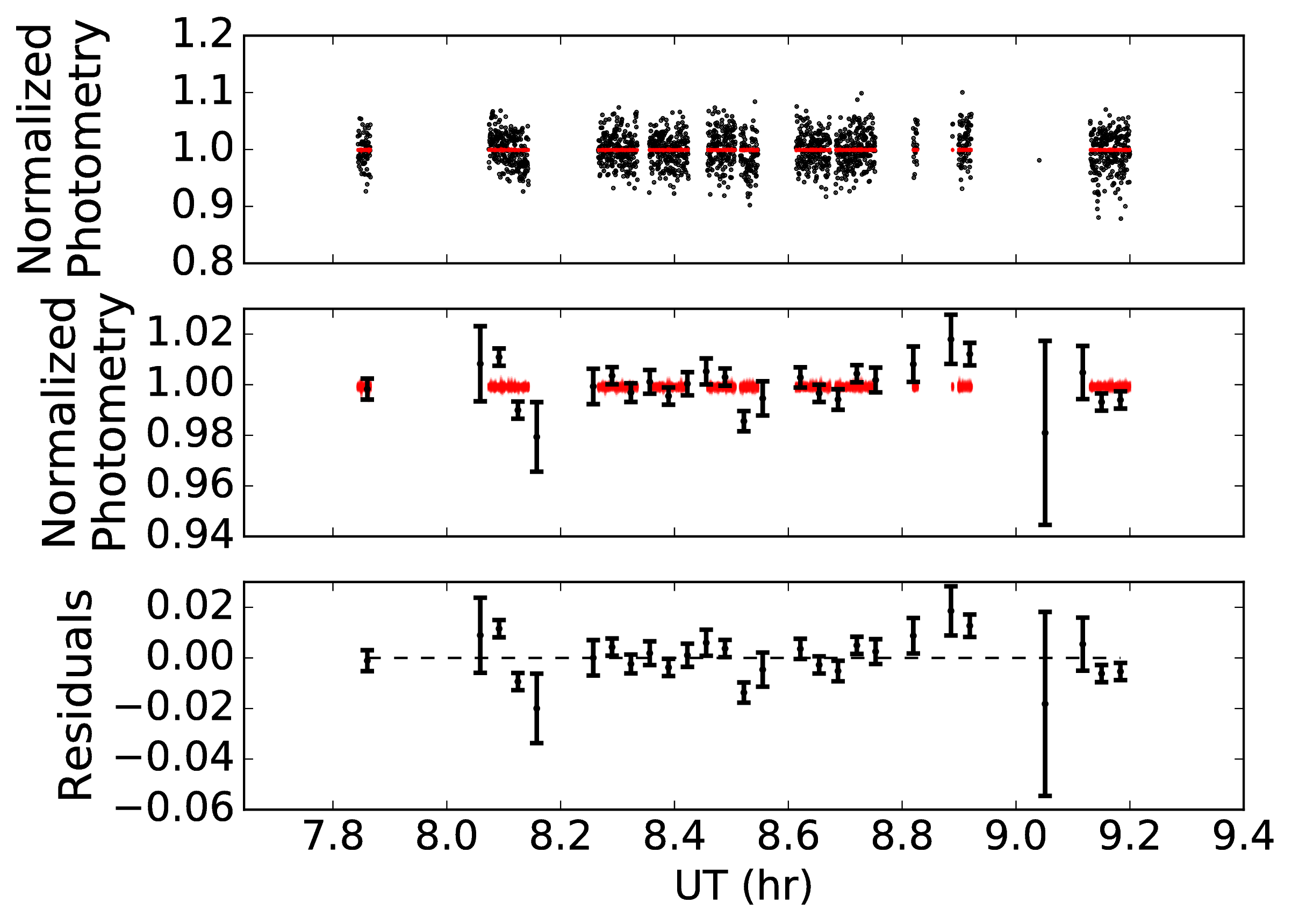}
\caption{Plots corresponding to the constant target dataset, with systematics terms indicated above the plots. Top row: unbinned photometry in black, with model posteriors overlaid in red. Middle row: the same as the top row, but with a rescaled ordinate and photometry in bins of two minutes for display. Note that the binned data point with large error bars at UT $\sim$9.05 is for a bin containing a single data point. Bottom row: the residuals between the photometry and the model, in bins of two minutes for display. For this dataset, the parametric systematics terms with nonzero wind models do not reproduce the systematics.} 
\label{fig:oct2014MCMCsamples}
\end{figure*}

\begin{figure*}
\begin{centering}
\begin{minipage}{0.333\textwidth}
	\centering
	$\Phi=0$
\end{minipage}
\begin{minipage}{0.333\textwidth}
	\centering
	$\Phi \propto \hat{v_{w}}\cdot \hat{B}$ 
\end{minipage}
\begin{minipage}{0.333\textwidth}
	\centering
	\hspace{-0.1in}
	$\Phi \propto \overrightarrow{v_{w}}\cdot \hat{B}$
\end{minipage}
\vspace{-0.1in}
\end{centering}\\
\includegraphics[width=6cm]{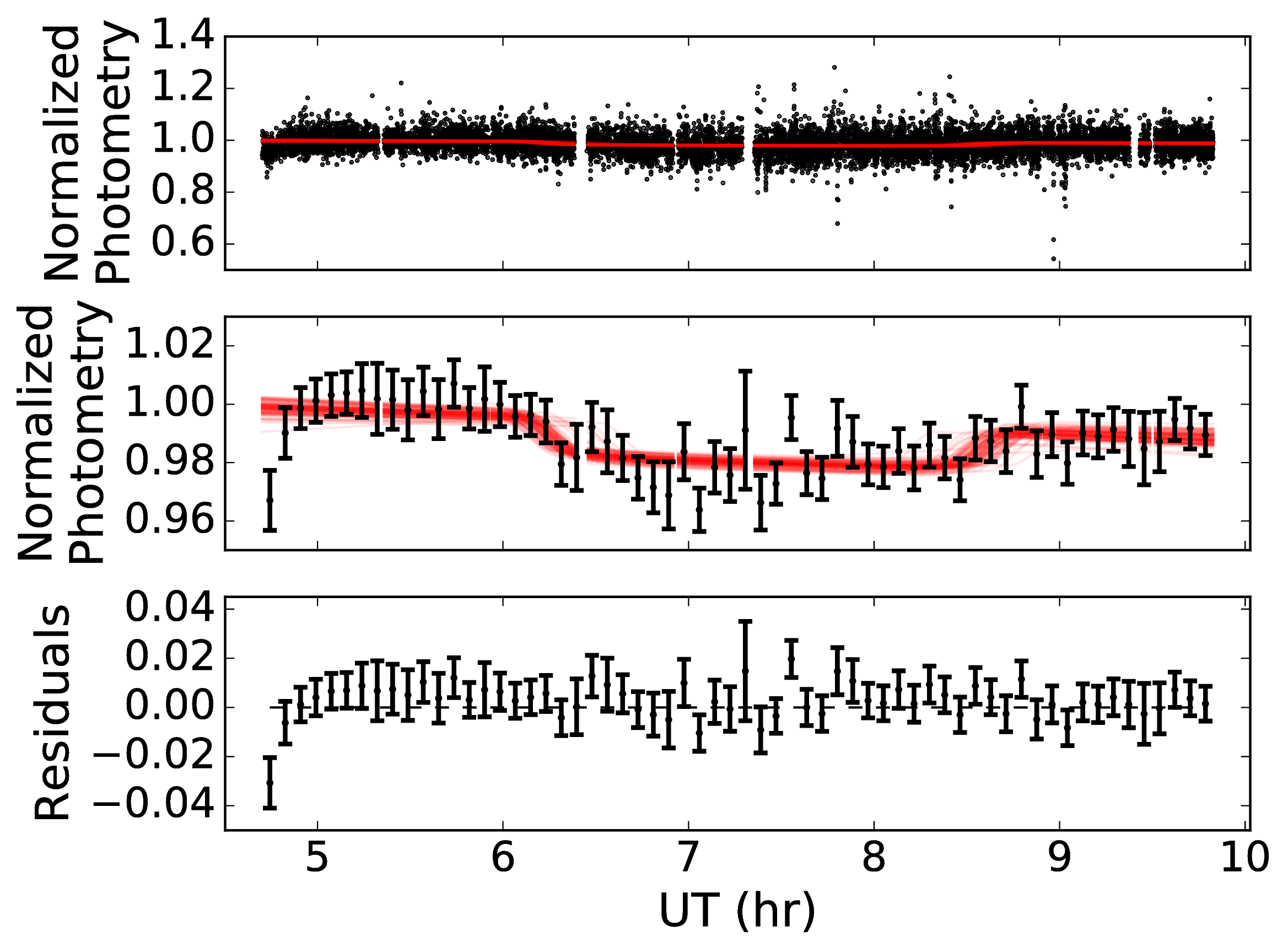}
\includegraphics[width=6cm]{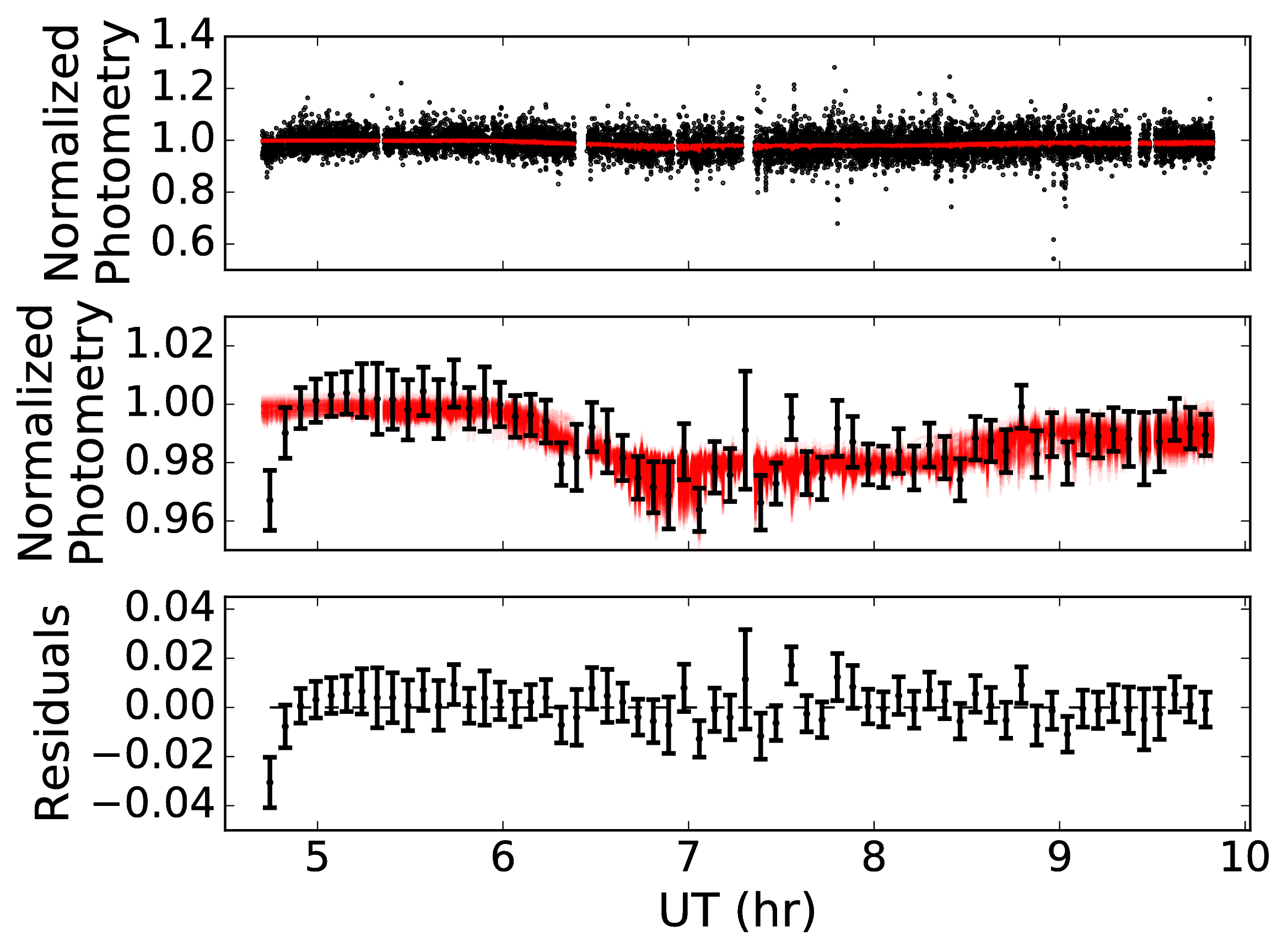}
\includegraphics[width=6cm]{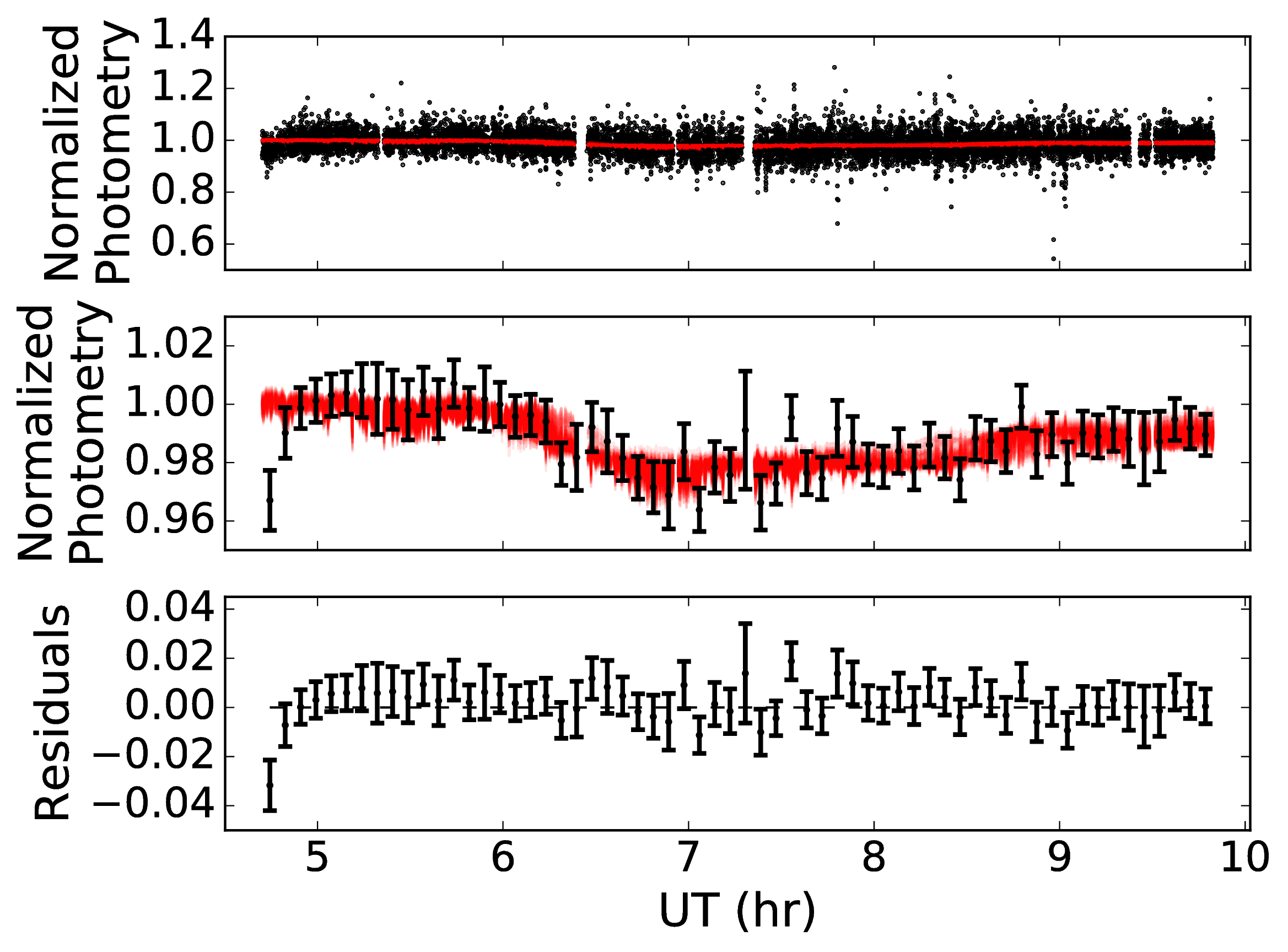}
\caption{The same as in Fig. \ref{fig:oct2014MCMCsamples}, for the primary transit dataset. Here the photometry is shown in bins of five minutes. This dataset exhibits possible wind-related systematics, including those present during the transit itself. If left uncorrected, such systematics can have a pernicious effect on the measured $R_{P}/R_{S}$.} 
\label{fig:feb2015MCMCsamples}
\end{figure*}

\begin{figure*}
\begin{centering}
\begin{minipage}{0.333\textwidth}
	\centering
	$\Phi=0$
\end{minipage}
\begin{minipage}{0.333\textwidth}
	\centering
	$\Phi \propto \hat{v_{w}}\cdot \hat{B}$ 
\end{minipage}
\begin{minipage}{0.333\textwidth}
	\centering
	\hspace{-0.1in}
	$\Phi \propto \overrightarrow{v_{w}}\cdot \hat{B}$
\end{minipage}
\vspace{-0.1in}
\end{centering}\\
\includegraphics[width=6cm]{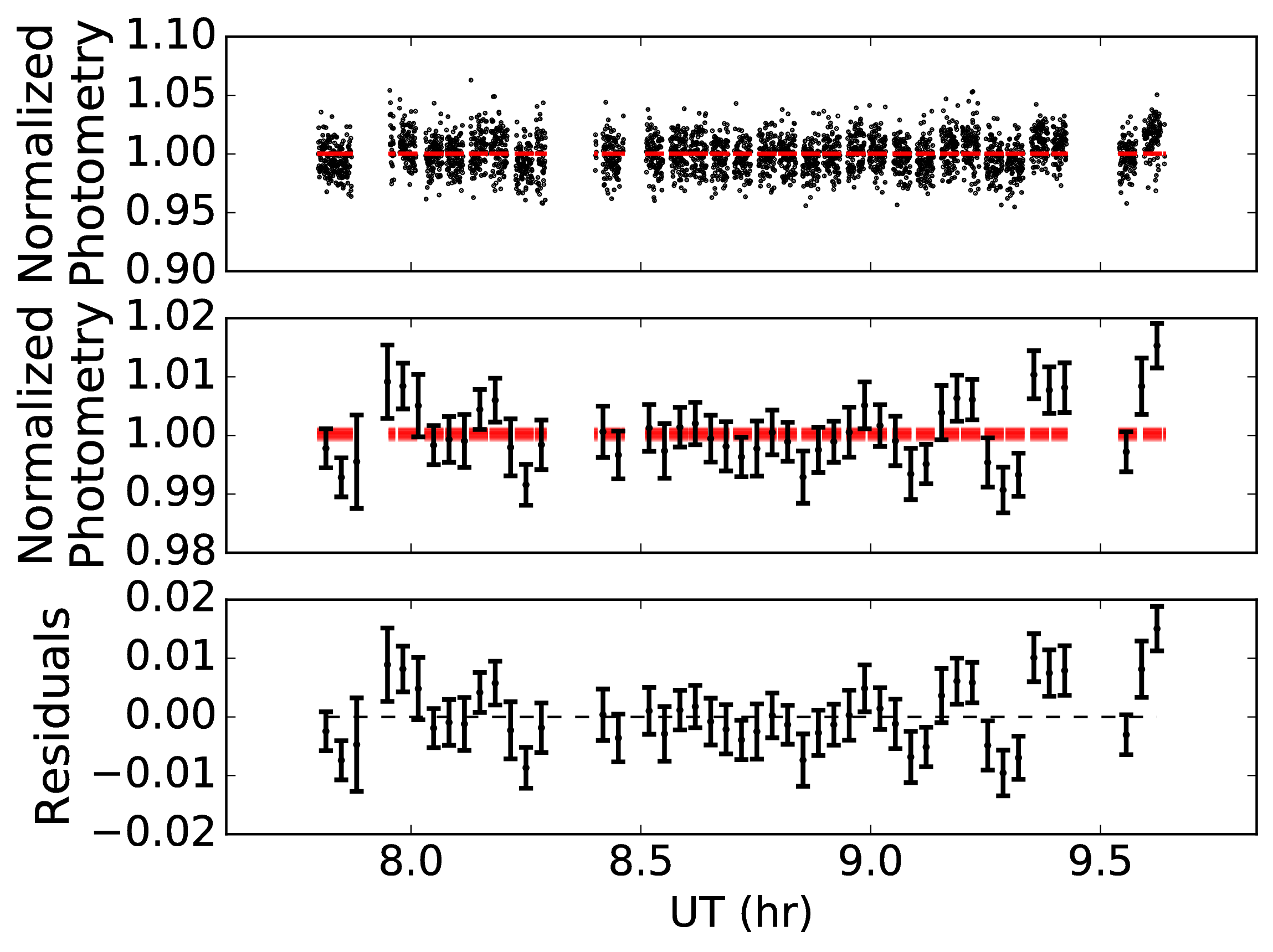}
\includegraphics[width=6cm]{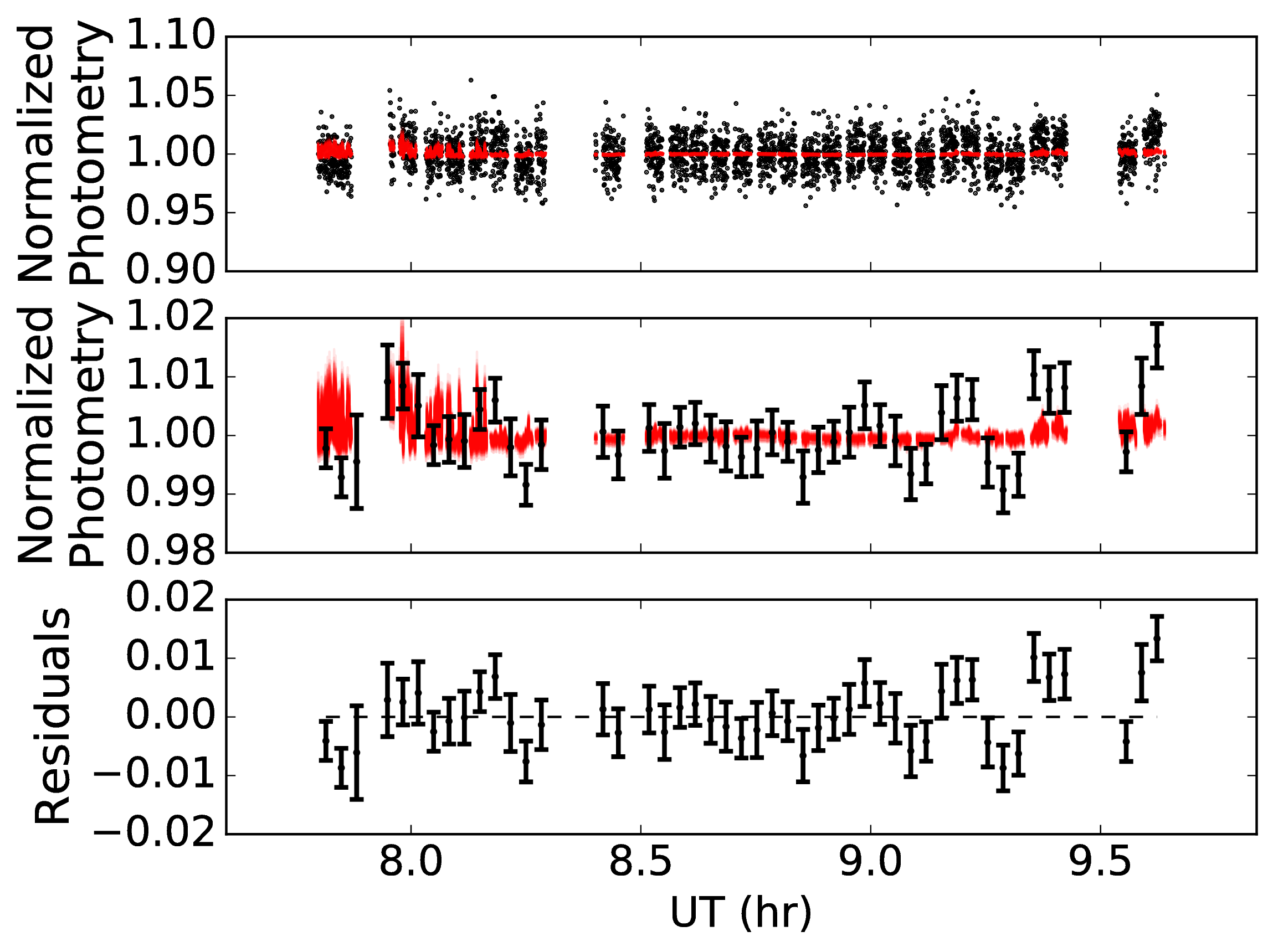}
\includegraphics[width=6cm]{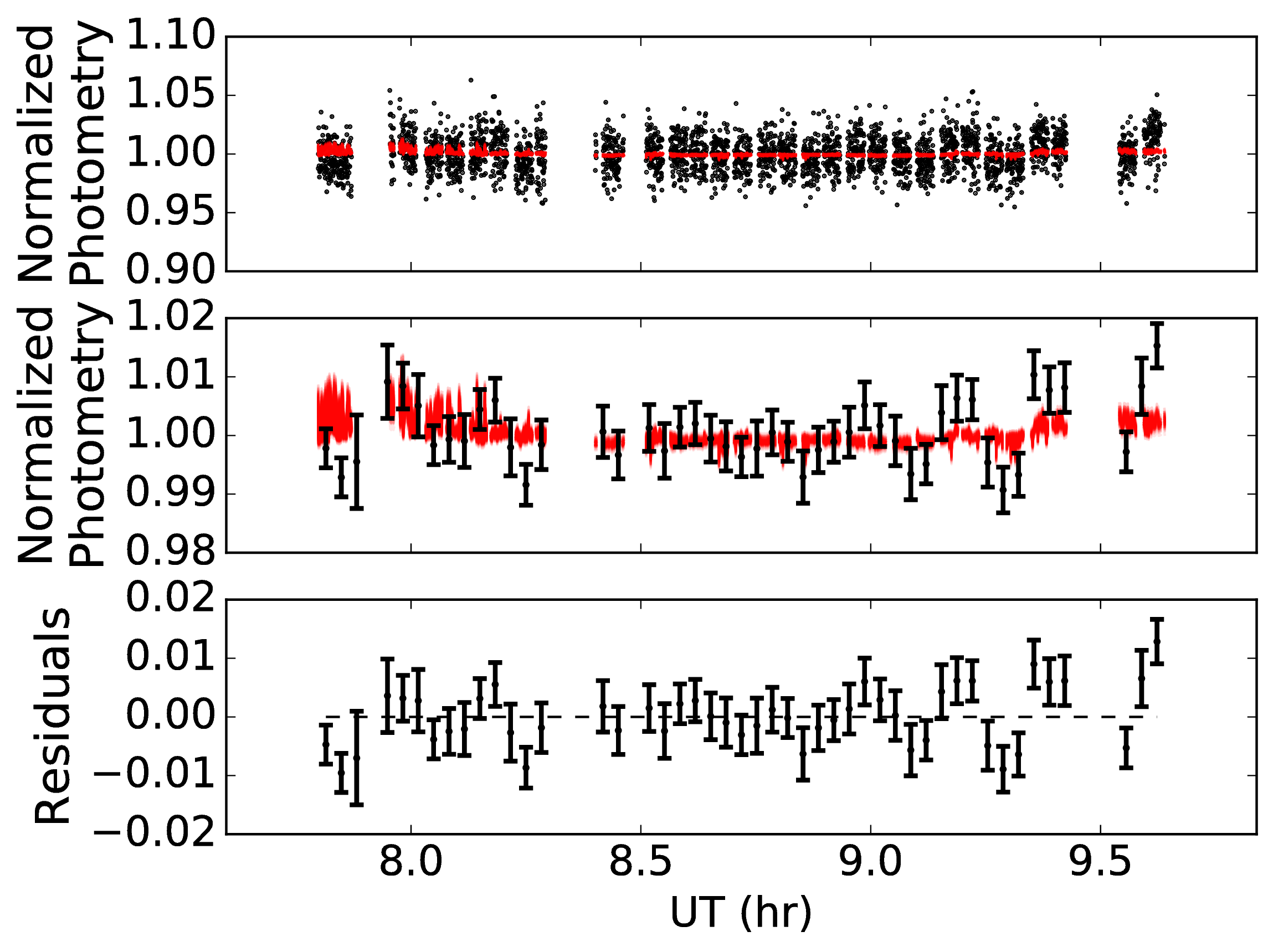}
\caption{The same as in Fig. \ref{fig:oct2014MCMCsamples}, for the secondary eclipse dataset. Here the photometry is shown in bins of two minutes. There appears to be a correlation between the systematics and wind across the apertures at UT $\lesssim$8.2, but other systematics appear to be dominant afterwards.} 
\label{fig:june2015MCMCsamples}
\end{figure*}

Figs. \ref{fig:oct2014MCMCsamples} to \ref{fig:june2015MCMCsamples} show comparisons of our photometry with samples of the MCMC posteriors of the models. In Fig. \ref{fig:oct2014MCMCsamples}, there is qualitatively little or no improvement based on the different systematics models. In Fig. \ref{fig:feb2015MCMCsamples}, inclusion of the effects of wind can be seen to reproduce the dominant systematics in the primary transit dataset. In Fig. \ref{fig:june2015MCMCsamples}, we see that wind effects help to explain the variations seen in the early part (UT $\lesssim8.2$) of the secondary eclipse observation, but not the apparently nod-related systematics in the second half.

\begin{figure}
\includegraphics[height=3.2cm]{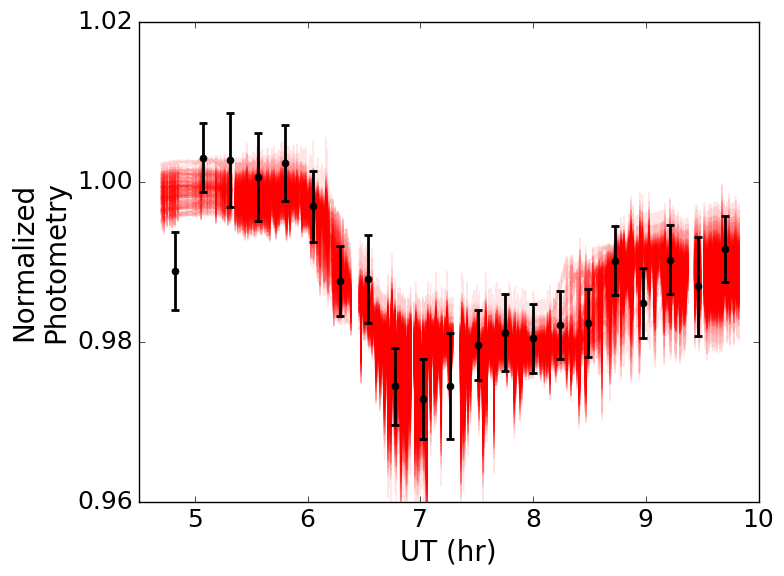}
\includegraphics[height=3.2cm, trim={2cm, 0cm, 0cm, 0cm}, clip=true]{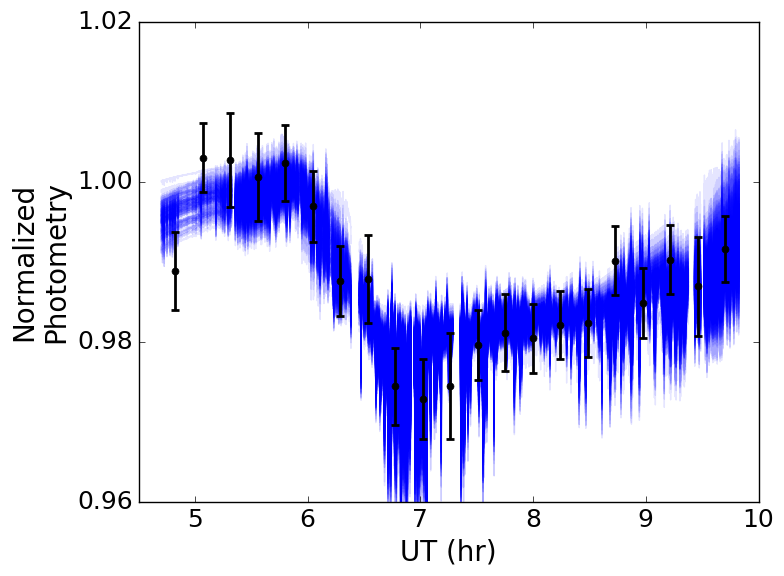} 
\caption{Left: A hundred posterior samples of the XO-2Nb primary transit (red) based on the systematics term $\Phi \propto \hat{v_{w}}\cdot \hat{B}$, showing the dominant mode ($t_{TOC}<7.8$) only. (See Fig. \ref{fig:feb2015corner}.) Empirical photometry is binned by 15 minutes and is overlaid with error bars. Right: the same, for the minor mode ($t_{TOC}>7.8$) only. It can be seen that the minor mode arises due to poor constraint of the transit egress.} 
\label{fig:twoModes}
\end{figure}

Fig. \ref{fig:feb2015corner} shows the MCMC posteriors of the parameters for the primary transit, using $\Phi \propto \hat{v_{w}}\cdot \hat{B}$. There are smaller modes visible in some of the 1D and 2D parameter space histograms. These smaller modes make up one distinct mode in multidimensional space. As seen in Fig. \ref{fig:twoModes}, this mode originates from the fact that the egress of the planet is near the very end of the observation, and that the egress and post-egress baseline are therefore poorly constrained. The histograms of the case $\Phi \propto \overrightarrow{v_{w}}\cdot \hat{B}$ are very similar to Fig. \ref{fig:feb2015corner}, except that the the minor mode is slightly larger. 

Following the fitting of the flux models, values of $\sigma_{1}$ and $\bar{\beta}$ were found from the photometric residuals with the models using the median coefficients from the posteriors. These are tabulated in Table \ref{table:redNoiseResults}. We quantified the appropriateness of the applied models by calculating the Akaike Information Criterion \citep{akaike1974new}, which in this case is

\begin{equation}
AIC = \chi^{2}+2k
\end{equation}

\noindent
where $k$ is the number of free parameters to be fit. We also find the Bayesian Information Criterion \citep{schwarz1978estimating}, or

\begin{equation}
BIC = \chi^{2}+kln(N)
\end{equation}

\noindent
where $N$ is the number of data points. The AIC and BIC have terms which penalize additional model parameters so as to avoid overfitting. When choosing among models which fulfill assumptions of the AIC or BIC, the one with the smallest AIC or BIC is considered the most appropriate to the data. This distinction, however, may or may not be significant. The significance of a model's appropriateness to the data depends on the difference $\Delta$AIC or $\Delta$BIC between that model and another model being compared. Values of $\Delta$AIC or $\Delta$BIC of $\sim2-6$ are considered to be `positive evidence' that the model with the minimum AIC or BIC is the most appropriate, $\sim6-10$ is `strong evidence', and $\gtrsim10$ is `very strong'. We tabulate values between the three systematics models in Table \ref{tab:goodnessOfFit}. 

In the case of the constant target dataset, the MCMC fit that uses $\Phi=0$ is favored by both the AIC and BIC (Table \ref{table:goodnessOfFit}). However, the AIC strongly favors the primary transit and secondary eclipse models with $\Phi\neq0$. According to the AIC, the primary transit data is best modeled using $\Phi \propto \hat{v_{w}}\cdot \hat{B}$, and the secondary eclipse data can either be modeled by the systematics terms $\Phi \propto \hat{v_{w}}\cdot \hat{B}$ or $\Phi\propto\overrightarrow{v_{w}}\cdot \hat{B}$. For the constant target and secondary eclipse datasets, the BIC tends to favor the models with $\Phi=0$. This is a result of the heavier penalization of additional parameters by the BIC. However, in the case of the primary transit, the BIC does not indicate positive evidence that either the $\Phi=0$ or $\Phi \propto \hat{v_{w}}\cdot \hat{B}$ is better than the other.

\floattable
\begin{deluxetable}{c|l|ccc}
\tablewidth{200pt}
\tabletypesize{\scriptsize}
\tablecaption{MCMC Model Goodness-of-Fit$^{a}$\label{tab:goodnessOfFit}}
\label{table:goodnessOfFit}
\tablehead{
\colhead{Dataset} & 
\colhead{Fit} & 
\colhead{$\chi^{2}_{\nu}$} &
\colhead{$\Delta AIC$} &
\colhead{$\Delta BIC$}
}
\startdata
\multirow{3}{*}{Constant targets}
&  $\Phi=0$ 	& 0.68  	& --  	& --\\
&  $\Phi\propto\hat{v_{w}}\cdot \hat{B}$ 	& 0.68  	& 3.2  	& 14.2\\
&  $\Phi\propto\overrightarrow{v_{w}}\cdot \hat{B}$ 	& 0.68  	& 4.0  	& 15.0\\
\hline
\multirow{3}{*}{XO-2Nb transit}
&  $\Phi=0$ 	& 0.16 	 	& 6.4 	 	& --\\
&  $\Phi\propto\hat{v_{w}}\cdot \hat{B}$ 	& 0.16 	 	& --		 	& 1.0\\
&  $\Phi\propto\overrightarrow{v_{w}}\cdot \hat{B}$ 	& 0.16 	 	& 4.3 	 	& 5.3\\
\hline
\multirow{3}{*}{HD 189733b sec. eclipse}
&  $\Phi=0$ 	& 0.22 	 	& 6.7 	 	& --\\
&  $\Phi\propto\hat{v_{w}}\cdot \hat{B}$ 	& 0.21 	 	& -- 	 	& 5.3\\
&  $\Phi\propto\overrightarrow{v_{w}}\cdot \hat{B}$ 	& 0.21 	 	& 0.4 	 	& 5.8\\
\enddata
\tablenotetext{a}{Within each dataset, the three systematics models are compared with each other by finding $\Delta$AIC and $\Delta$BIC with respect to the model with the minimum AIC and BIC. (Entries of `--' exist where the model with the minimum value would be compared with itself.) For example, the entry of $\Delta$AIC$=6.4$ for the $\Phi=0$ systematics term in the primary transit dataset indicates `strong evidence' that the $\Phi\propto\hat{v_{w}}\cdot \hat{B}$ systematics term is more appropriate to the data than $\Phi=0$. (See Sec. \ref{sec:results}.)}
\end{deluxetable}

The BIC is preferable if the true model is among those tested, in which case it is guaranteed to choose the correct model as the number of data points go to infinity. Since the BIC does not strongly and exclusively favor either of our wind models, it may be that different and more elaborate systematics models involving wind (or other parameters at higher altitudes) would produce better fits to the data. BIC also assumes that the $N$ data points are truly independent \citep{jones2011bayesian}. However, Figs. \ref{fig:oct2014MCMCsamples}, \ref{fig:feb2015MCMCsamples}, and \ref{fig:june2015MCMCsamples} suggest that the photometric residuals still contain correlated noise not captured by the parametric model. Neither are the MCMC posteriors (such as in Fig. \ref{fig:feb2015corner}) gaussian 
\citep{kuha2004aic,liddle2007information,vrieze2012model}. The AIC is thus probably more applicable in our analysis for finding an ersatz for a more complicated model of the systematics.

\newpage

\newpage


Using a systematics term of $\Phi \propto \hat{v_{w}}\cdot \hat{B}$, we found a transit depth in the normalized, relative photometry of $R_{P}/R_{S}=\sqrt{\Delta F}=0.1036^{+0.0080}_{-0.0072}$ (Table \ref{tab:transitResults}). Using a systematics term of $\Phi \propto \overrightarrow{v_{w}}\cdot \hat{B}$, we find $R_{P}/R_{S}=0.1034^{+0.0092}_{-0.0081}$ but with poorer constraints on the total transit time $T_{tot}$. Both results for $R_{P}/R_{S}$ are consistent with the same transit event as observed by \citet{zellem2015xo} in the Bessel $U$ and Harris $B$ bands using the University of Arizona's 61-inch (1.55 m) Kuiper Telescope on Mt. Bigelow (also Table \ref{tab:transitResults}). Application of the preceding analysis to the photometry from the science target alone (with $\Phi=0$, since there is no baseline for wind to cross) yields a clearly erroneous $R_{P}/R_{S}=0.1980^{+0.0055}_{-0.0049}$. 

To test the influence of the prior on the transit depth, we reran the  MCMC for $\Phi \propto \hat{v_{w}}\cdot \hat{B}$ with a gaussian prior on $R_{P}/R_{S}$ twice as wide as in Table \ref{table:priors}. This resulted in a significantly larger minor mode and changed the transit depth to $R_{P}/R_{S}=0.110^{+0.032}_{-0.016}$. This illustrates the importance of obtaining a large amount of pre- and post-egress baseline, especially in the absence of a plausible narrow prior.

In the secondary eclipse observation, the scatter before the change in the state of the right adaptive secondary mirror was $\sigma_{1}=0.015$. This was greater than the expected depth of the transit, and the relative flux changed by $\sim$3\% after the change in state of the adaptive secondary mirror. Therefore, the photometry after the adaptive secondary mirror state change was masked.

\begin{figure}
\centering
\includegraphics[width=1.0\linewidth]{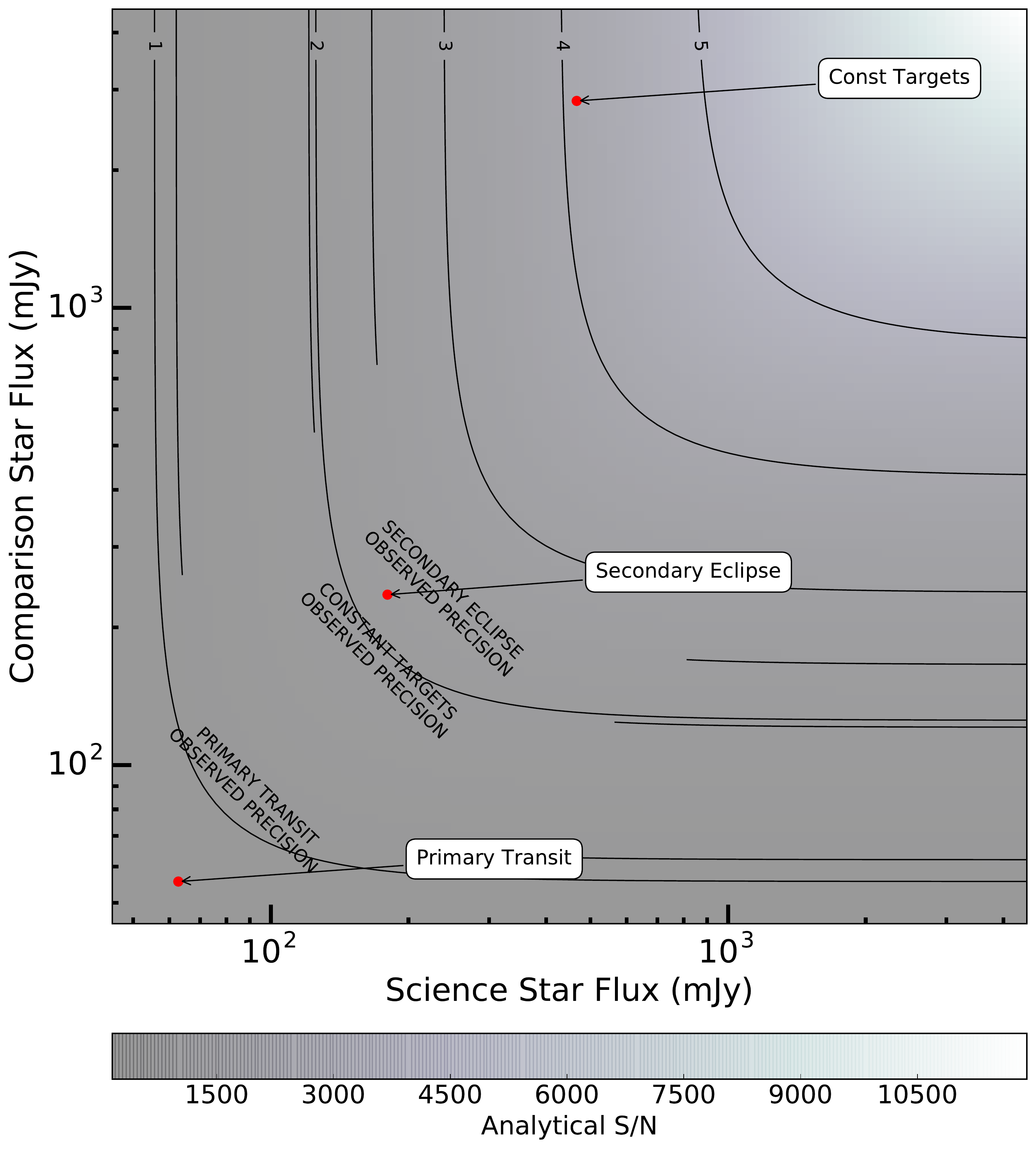} \\
\caption{Expected signal-to-noise, as described in Sec. \ref{sec:discussion}. The background level, photometry aperture, and photometric bin sizes are chosen to replicate the primary transit dataset. The contours either indicate 1.) minimum levels of signal-to-noise required for measuring the amplitudes of different science goals, which are labeled with numbers, or 2.) the observed precisions of the three datasets, which are labeled with text. Keys to the numbers are in Table \ref{table:ston_numberKey}. Red points indicate the true brightnesses of the stellar pairs in our datasets. Their distance from the empirical contours are due to systematics not accounted for in the model described in Sec.  \ref{sec:discussion}, and we emphasize that these plots should only be taken to provide very rough predictive estimates.} 
\label{fig:stonplots}
\end{figure}

\begin{table}
\begin{flushleft}
\caption{Science requirements key}
\label{table:ston_numberKey}
\begin{tabular}{ p{.05\linewidth} p{.35\linewidth} p{.25\linewidth} p{.17\linewidth} }
\hline
\textit{Num}     &     \textit{Description}  & \textit{Prototype$^{a}$}            &               \textit{S/N}    \\
\hline
$-$           &               Large brown dwarf variability            &       SIMP 0136          & $\sim$60 \\ 
\hline
   $-$        &               Deep primary transit             &       HATS-6b          & $\sim$93 \\ 
\hline
\textbf{$1$}          &               Primary transit           &       XO-2Nb            &       $\sim$278 \\ 
\hline
        &               \textit{Observed primary transit precision}          &   $-$ &   $\sim$310  \\ 
\hline
          &               \textit{Observed constant targets precision}            &  $-$ &  $\sim$600   \\ 
\hline
\textbf{$2$}          &               Deep secondary eclipse, 3.6 $\mu$m           &       WASP-19b            &   $\sim$621      \\ 
\hline
         &               \textit{Observed secondary eclipse precision}             &   $-$ &   $\sim$820  \\   
\hline
\textbf{$3$}          &               Secondary eclipse            &       HD 189733b            &    $\sim$1172 \\ 
\hline
\textbf{$4$}          &               Occultation Spectroscopy            &       (WASP-19b)            &   $\sim$2070 \\ 
\hline
\textbf{$5$}          &               Occultation Spectroscopy            &       (HD 189733b)            &      $\sim$3906     \\ 
\hline
\textbf{$-$}          &               Transmission Spectroscopy            &       (HATS-6b)            &    $\sim3\times10^{4}$   \\ 
\hline
\\\end{tabular}      
\tablenotetext{a}{Prototypes are only meant to set the amplitude of variation, and do not take into account the brightness of any targets actually located at that point in the sky.}       
\noindent
\end{flushleft}
\end{table}

\section{Discussion} \label{sec:discussion}

Differential photometry is useful insofar as systematic noise is correlated between the photometries of different PSFs. Using `wall-eyed' pointing, we have shown that we can remove correlated noise with differential photometry. In addition, we have used a parametric noise model to obtain an accurate primary transit depth despite the persistence of uncorrelated residual noise. This suffices to achieve percent-level precision in time-series photometry for bright targets, with the possibility of obtaining better precision with longer observations. 

The LBT is a unique, infrared-optimized platform for carrying out these observations. Other large telescope twins (such as Keck or any VLT telescope pairs) have separations larger than that of the 14.4 m LBT aperture separation, in which case the degree of noise correlation between separately-observed PSFs will likely decrease. However, the precision of `wall-eyed' pointing will not outperform space-based observations at these wavelengths, such as with Spitzer/IRAC, which can already reach precisions on the order of 100 ppm for stars of brightnesses comparable to those we have used in this work \citep{todorov2009spitzer,ingalls2016repeatability}. 

Nevertheless, `wall-eyed' pointing expands the available infrastructure for performing differential photometry of northern targets,\footnote{For efforts to achieve thermal infrared differential photometry in the southern hemisphere, see \citet{kamizuka2014revised}} and partly circumvents the bottleneck imposed by the heavy oversubscription of space-based facilities. A ground-based facility like the LBT also allows instrument upgrades and filter changes as needs evolve. 

`Wall-eyed' pointing would be of particular benefit to observations at low galactic latitudes, where it is more likely to find a comparison star within 2 amin. From the 2MASS dataset \citep{skrutskie2006two}, densities of stars in the galactic plane with $K_{S}\leq11$ range between the equivalent of 8.73 stars within circular cones 2 amin in diameter towards the Galactic center, and 0.35 stars toward the anticenter. These densities decrease with galactic latitude, down to 0.03 stars per cone straight out of the galactic plane \citep{polido2013galaxy}. 


Given a suitable target pair, the decorrelation in the signals along the two lines-of-sight tends to be dominated by environmental effects. We find a possible correlation between systematics in at least one of our datasets with a changing gradient of effective transmissivity across the two lines-of-sight. Whether this is due to true atmospheric transmissivity changes is unknown. The time-varying number of photons that actually reach the primary mirror and are corralled into the photometry apertures may be caused by wind-borne `blobs' of turbulence that lead to time-dependency in the differential quality of the AO correction. An alternative theory is that we are indeed seeing the effects of wind-borne water vapor which causes a true differential transmissivity along both lines of sight.

Based on the optically-measured transit depth of XO-2Nb and any given differential transmissivity $T_{s}/T_{c}$ between the science and comparison stars outside the transit, the ratio $T_{s}/T_{c}$ need only decrease by $\sim2\times10^{-3}$ during the transit to mimic the $R_{P}/R_{S}=0.1132$ we found using the model with $\Phi=0$. If such a difference is being caused by precipitable water vapor changes and not turbulent blobs, we can see how large of a PWV change would be necessary by integrating over interpolated Gemini transmission spectra. If both LBT apertures `see' an effective PWV of 1.6 mm at an airmass of 1.0 outside the transit, but the aperture pointing towards the science target `sees' a PWV of 1.7 mm during the transit, this will incur a change of $\sim4\times10^{-3}$ in $T_{s}/T_{c}$. At an airmass of 1.5, this becomes $\sim5\times10^{-3}$. 

Interpolations of transmission spectra based on the recorded airmass and SMT PWV values show that the general level of transmission changed considerably during this observation, over a total range of $>$5\% \citep{spalding2016infrared}. However, proof of the culpability of wind-borne PWV variations would require a better understanding of the connection between ground-level windspeed, changing wind profiles at higher altitudes, and the time-dependent behavior of differential PWV values along the lines-of-sight. 

Even if the nature of the systematics remains unproven, use of a quantity $\hat{v_{w}}\cdot \hat{B}$ or $\overrightarrow{v_{w}}\cdot \hat{B}$ can be useful for removing some of the systematics in the differential photometry. Both terms yielded essentially the same planet-to-star radius (see Sec. \ref{sec:results}) in our primary transit dataset, so we cannot rule out one wind model in favor of the other. These terms also capture some of the systematics seen in the early part of the secondary eclipse observation, but are not seen to reproduce the systematics in the constant target data. Thus we offer these models as a \textit{possible} tool for removing residual systematics, but they do not capture the full systematics. 

The precision of the found $R_{P}/R_{S}$ is equivalent to a flux precision of $\sim3\times10^{-3}$, which is consistent with the binned scatter results in Table \ref{table:redNoiseResults}.  In the same Table we see that the binned scatter results of the constant target and secondary eclipse datasets are even lower, but this may be simply due to the large bin sizes relative to the datasets. We stress that there is still red noise present in the residuals (i.e., $\beta>1$). The found standard deviations are meant to give a general idea of the scatter present in the residuals, but are not meant to imply that the residuals behave like Gaussian white noise. Therefore we consider the obtainable precision of `wall-eyed' pointing to be on the percent level, or better if the noise can be well characterized. 


We make coarse predictions of `wall-eyed' pointing's scientific potential by considering the scatter in the differential flux $F_{s}/F_{c}$ for different target brightnesses, and make conservative estimates by treating the two fluxes as uncorrelated. Assuming the standard deviations are small, independent, and Gaussian, we propagate errors and include red noise $\beta$ and binning parameters from the primary transit dataset to find the expected scatter $\sigma_{Nmax}'$ (as defined in Table \ref{table:redNoiseResults}) in the normalized $F_{s}/F_{c}$. We take the inverse of $\sigma_{Nmax}'$ to find the S/N landscape. This is plotted in Fig. \ref{fig:stonplots}, with contours to indicate the empirical scatter in our three datasets, as well as threshold signals-to-noise for different science applications. These thresholds are found by taking the inverses of the amplitudes of variations and multiplying them by a factor of three (except for occultation spectroscopy, where the inverse of the secondary eclipse variation has been multiplied by ten \citep{crossfield2015}).

From Fig. \ref{fig:stonplots} and Table \ref{table:ston_numberKey}, one can see that many exoplanet-related science signals are outside the range of what has been obtained in this analysis. However, this provides impetus for improvement by taking additional observations with the following general guidelines in mind. The verticality of the contours in Fig. \ref{fig:stonplots} for bright comparison stars shows that ever brighter comparison stars provide less additional benefit. Based on the nonlinear response of LMIRcam at high ADU counts, care should be taken to linearize the response if the two target brightnesses are more than one or two magnitudes apart. Also, since our data was so strongly affected by environmental parameters, future observations should record all potentially relevant telemetry, and to observe in low-wind conditions if possible. Of course, future analyses of the noise based either on parametric models (like Eqn. \ref{eqn:systemEqn}) or nonparametric models (such as gaussian processes) will benefit from observations which are over longer time baselines for better sampling of the systematics.

For observations so strongly affected by high sky backgrounds and water vapor variations, adaptive optics has the benefit of minimizing the sky photon noise, but its complexity introduces an additional element of risk. `Wall-eyed' pointing fails if one of the adaptive secondary mirrors fails during a critical part of the observation, because changes in the AO correction before and after such an event may affect the measured flux. The likelihood of triggering a mirror to rest in a safe mode depends on software issues, the history of the mirror, and environmental conditions. The latter provides another reason to observe in good conditions. 

If the mirror safe event during our HD 189733b secondary eclipse observation had not occurred, it is still doubtful that the signal would have been reliably detected in this observation, given the $\sim$15 minutes of baseline before the start of the eclipse. The scatter found in this dataset (Table \ref{table:redNoiseResults}) suggests that secondary eclipses at least as deep may be possible to detect (Table \ref{table:ston_numberKey}), though such an observation should include much more baseline outside of the event.

\section{Summary} \label{sec:summary}

We have demonstrated the possibility of obtaining time-series photometry in $L_{S}$-band with a new `wall-eyed' pointing mode at the Large Binocular Telescope. This observing mode is applicable to science targets with comparison stars $\lesssim$2 amin away, and may be the only way of simultaneously obtaining a comparison star for observations at these wavelengths. Differential photometry removes noise from the science signal which is correlated with noise in the comparison star signal. This correlation is only partial, however, and a parametric systematics model based on the roof-level wind vector may be useful for removing some of the residual noise in the differential signal. At this stage, the precision of photometry can be expected to be on the level of a percent, though the possibility remains for improvement with better characterization of the noise.

\acknowledgments
\section{Acknowledgments} \label{sec:acknowledgements}

We thank the anonymous reviewers for comments that improved the quality of this paper. This work was also made possible through discussions and exchanges with Yifan Zhou \begin{CJK*}{UTF8}{gbsn}(周一凡),\end{CJK*} Rob Zellem, Katie Morzinski, Denis Defr\`{e}re,  Jordan Stone, Ian Crossfield, D\'{a}niel Apai, Travis Barman, Dmitrios Psaltis, Jarron Leisenring, Andrew Youdin, Josh Winn, Kaitlin Kratter, and Daniel Foreman-Mackey. In addition, Joanna Bulger and Steph Sallum assisted during the Feb 2015 observations, and Yifan Zhou assisted during the June 2015 observations. Zhou also kindly allowed the use of his centroiding algorithm. 

The Large Binocular Telescope Interferometer is funded by the National Aeronautics and Space Administration within the framework of the Exoplanet Exploration Program. The LBT is an international collaboration among institutions in the United States, Italy and Germany. LBT Corporation partners are: The University of Arizona on behalf of the Arizona Board of Regents; Istituto Nazionale di Astrofisica, Italy; LBT Beteiligungsgesellschaft, Germany, representing the Max-Planck Society, The Leibniz Institute for Astrophysics Potsdam, and Heidelberg University; The Ohio State University, and The Research Corporation, on behalf of The University of Notre Dame, University of Minnesota and University of Virginia.

High-resolution atmospheric transmission and emission data is based on observations obtained at the Gemini Observatory, which is operated by the Association of Universities for Research in Astronomy, Inc., under a cooperative agreement with the NSF on behalf of the Gemini partnership: the National Science Foundation (United States), the National Research Council (Canada), CONICYT (Chile), Ministerio de Ciencia, Tecnolog\'{i}a e Innovaci\'{o}n Productiva (Argentina), and Minist\'{e}rio da Ci\^{e}ncia, Tecnologia e Inova\c{c}\~{a}o (Brazil). 

This publication makes use of data products from the Two Micron All Sky Survey, which is a joint project of the University of Massachusetts and the Infrared Processing and Analysis Center/California Institute of Technology, funded by the National Aeronautics and Space Administration and the National Science Foundation. This research has also made use of the SIMBAD database, operated at CDS, Strasbourg, France. 

This work has made use of data from the European Space Agency (ESA)
mission {\it Gaia} (\url{https://www.cosmos.esa.int/gaia}), processed by
the {\it Gaia} Data Processing and Analysis Consortium (DPAC,
\url{https://www.cosmos.esa.int/web/gaia/dpac/consortium}). Funding
for the DPAC has been provided by national institutions, in particular
the institutions participating in the {\it Gaia} Multilateral Agreement.

Some of the MCMC computations were conducted on CyVerse Atmosphere cyberinfrastructure, which is supported by the National Science Foundation under Award Numbers DBI-0735191 and DBI-1265383. URL: \url{www.cyverse.org}

Corner plots were made with the Python \texttt{corner} package \citep{corner}.

\vspace{5mm}
\facilities{LBT}

\software{IDL, Python, \texttt{emcee}}

\bibliographystyle{apj}
\bibliography{bibliog}

\begin{thebibliography}{}
\expandafter\ifx\csname natexlab\endcsname\relax\def\natexlab#1{#1}\fi

\bibitem[{Akaike(1974)}]{akaike1974new}
Akaike, H. 1974, IEEE Transactions on Automatic Control, 19, 716

\bibitem[{Birkby {et~al.}(2013)Birkby, de~Kok, Brogi, de~Mooij, Schwarz,
  Albrecht, \& Snellen}]{birkby2013detection}
Birkby, J., de~Kok, R., Brogi, M., {et~al.} 2013, MNRASL, 436, L35

\bibitem[{Brown {et~al.}(2016)Brown, Vallenari, Prusti, De~Bruijne, Mignard,
  Drimmel, Babusiaux, Bailer-Jones, Bastian, Biermann,
  {et~al.}}]{brown2016gaia}
Brown, A.~G., Vallenari, A., Prusti, T., {et~al.} 2016, A\&A, 595, A2

\bibitem[{Buenzli {et~al.}(2012)Buenzli, Apai, Morley, Flateau, Showman,
  Burrows, Marley, Lewis, \& Reid}]{buenzli2012vertical}
Buenzli, E., Apai, D., Morley, C.~V., {et~al.} 2012, ApJL, 760, L31

\bibitem[{Burke {et~al.}(2007)Burke, McCullough, Valenti, Johns-Krull, Janes,
  Heasley, Summers, Stys, Bissinger, Fleenor, {et~al.}}]{burke2007xo}
Burke, C.~J., McCullough, P., Valenti, J.~A., {et~al.} 2007, ApJ, 671, 2115

\bibitem[{Carter \& Winn(2009)}]{carter2009parameter}
Carter, J.~A., \& Winn, J.~N. 2009, ApJ, 704, 51

\bibitem[{Chamberlin \& Bally(1994)}]{chamberlin1994225}
Chamberlin, R.~A., \& Bally, J. 1994, Applied Optics, 33, 1095

\bibitem[{Charbonneau {et~al.}(2008)Charbonneau, Knutson, Barman, Allen, Mayor,
  Megeath, Queloz, \& Udry}]{charbonneau2008broadband}
Charbonneau, D., Knutson, H.~A., Barman, T., {et~al.} 2008, ApJ, 686, 1341

\bibitem[{Croll {et~al.}(2015)Croll, Albert, Jayawardhana, Cushing, Moutou,
  Lafreniere, Johnson, Bonomo, Deleuil, \& Fortney}]{croll2015near}
Croll, B., Albert, L., Jayawardhana, R., {et~al.} 2015, ApJ, 802, 28

\bibitem[{Crossfield(2015)}]{crossfield2015}
Crossfield, I. 2015, PASP, 127, 941

\bibitem[{Crouzet {et~al.}(2012)Crouzet, McCullough, Burke, \&
  Long}]{crouzet2012transmission}
Crouzet, N., McCullough, P.~R., Burke, C., \& Long, D. 2012, ApJ, 761, 7

\bibitem[{Crouzet {et~al.}(2014)Crouzet, McCullough, Deming, \&
  Madhusudhan}]{crouzet2014water}
Crouzet, N., McCullough, P.~R., Deming, D., \& Madhusudhan, N. 2014, ApJ, 795,
  166

\bibitem[{Cutri {et~al.}(2003)Cutri, Skrutskie, Van~Dyk, Beichman, Carpenter,
  Chester, Cambresy, Evans, Fowler, Gizis, {et~al.}}]{cutri2003vizier}
Cutri, R., Skrutskie, M., Van~Dyk, S., {et~al.} 2003, VizieR Online Data
  Catalog, 2246

\bibitem[{de~Kok {et~al.}(2013)de~Kok, Brogi, Snellen, Birkby, Albrecht, \&
  de~Mooij}]{de2013detection}
de~Kok, R.~J., Brogi, M., Snellen, I.~A., {et~al.} 2013, A\&A, 554, A82

\bibitem[{Defr{\`e}re {et~al.}(2016)Defr{\`e}re, Hinz, Mennesson, Hoffmann,
  Millan-Gabet, Skemer, Bailey, Danchi, Downey, \& Durney}]{defrere2016nulling}
Defr{\`e}re, D., Hinz, P., Mennesson, B., {et~al.} 2016, ApJ, 824, 66

\bibitem[{Deming {et~al.}(2007)Deming, Richardson, \& Harrington}]{deming20073}
Deming, D., Richardson, L.~J., \& Harrington, J. 2007, MNRAS, 378, 148

\bibitem[{D{\'e}sert {et~al.}(2009)D{\'e}sert, Des~Etangs, H{\'e}brard, Sing,
  Ehrenreich, Ferlet, \& Vidal-Madjar}]{desert2009search}
D{\'e}sert, J.-M., Des~Etangs, A.~L., H{\'e}brard, G., {et~al.} 2009, ApJ, 699,
  478

\bibitem[{Esposito {et~al.}(2010)Esposito, Riccardi, Fini, Puglisi, Pinna,
  Xompero, Briguglio, Quir{\'o}s-Pacheco, Stefanini, Guerra,
  {et~al.}}]{esposito2010first}
Esposito, S., Riccardi, A., Fini, L., {et~al.} 2010, in SPIE Proc, 773609

\bibitem[{Fernandez {et~al.}(2009)Fernandez, Holman, Winn, Torres, Shporer,
  Mazeh, Esquerdo, \& Everett}]{fernandez2009transit}
Fernandez, J.~M., Holman, M.~J., Winn, J.~N., {et~al.} 2009, AJ, 137, 4911

\bibitem[{Foreman-Mackey(2016)}]{corner}
Foreman-Mackey, D. 2016, JOSS, 24, doi:10.21105/joss.00024

\bibitem[{Foreman-Mackey {et~al.}(2013)Foreman-Mackey, Hogg, Lang, \&
  Goodman}]{foreman2013emcee}
Foreman-Mackey, D., Hogg, D.~W., Lang, D., \& Goodman, J. 2013, PASP, 125, 306

\bibitem[{Grillmair {et~al.}(2008)Grillmair, Burrows, Charbonneau, Armus,
  Stauffer, Meadows, van Cleve, von Braun, \& Levine}]{grillmair2008strong}
Grillmair, C.~J., Burrows, A., Charbonneau, D., {et~al.} 2008, Nature, 456, 767

\bibitem[{Henning \& Semenov(2013)}]{henning2013chemistry}
Henning, T., \& Semenov, D. 2013, Chemical Reviews, 113, 9016

\bibitem[{Hill(2010)}]{hill2010large}
Hill, J.~M. 2010, Applied Optics, 49, D115

\bibitem[{Hinz {et~al.}(2004)Hinz, Connors, McMahon, Cheng, Peng, Hoffmann,
  McCarthy~Jr, \& Angel}]{hinz2004large}
Hinz, P.~M., Connors, T., McMahon, T., {et~al.} 2004, in SPIE Proc, 787

\bibitem[{Hinz {et~al.}(2008)Hinz, Solheid, Durney, \& Hoffmann}]{hinz2008nic}
Hinz, P.~M., Solheid, E., Durney, O., \& Hoffmann, W.~F. 2008, in SPIE Proc,
  701339

\bibitem[{Hoffmann {et~al.}(2014)Hoffmann, Hinz, Defr{\`e}re, Leisenring,
  Skemer, Arbo, Montoya, \& Mennesson}]{hoffmann2014operation}
Hoffmann, W.~F., Hinz, P.~M., Defr{\`e}re, D., {et~al.} 2014, in SPIE Proc,
  91471O

\bibitem[{Howarth(2011)}]{howarth2011new}
Howarth, I.~D. 2011, MNRAS, 413, 1515

\bibitem[{Ingalls {et~al.}(2016)Ingalls, Krick, Carey, Stauffer, Lowrance,
  Grillmair, Buzasi, Deming, Diamond-Lowe, Evans,
  {et~al.}}]{ingalls2016repeatability}
Ingalls, J.~G., Krick, J., Carey, S., {et~al.} 2016, AJ, 152, 44

\bibitem[{Jones(2011)}]{jones2011bayesian}
Jones, R.~H. 2011, Statistics in medicine, 30, 3050

\bibitem[{Kamizuka {et~al.}(2014)Kamizuka, Miyata, Sako, Ohsawa, Asano,
  Uchiyama, Okada, Uchiyama, Nakamura, \& Sakon}]{kamizuka2014revised}
Kamizuka, T., Miyata, T., Sako, S., {et~al.} 2014, in SPIE Proc, 91473C

\bibitem[{Kammer {et~al.}(2015)Kammer, Knutson, Line, Fortney, Deming, Burrows,
  Cowan, Triaud, Agol, Desert, {et~al.}}]{kammer2015spitzer}
Kammer, J.~A., Knutson, H.~A., Line, M.~R., {et~al.} 2015, ApJ, 810, 118

\bibitem[{Kuha(2004)}]{kuha2004aic}
Kuha, J. 2004, Sociological methods \& research, 33, 188

\bibitem[{Lee {et~al.}(2012)Lee, Fletcher, \& Irwin}]{lee2012optimal}
Lee, J.-M., Fletcher, L.~N., \& Irwin, P.~G. 2012, MNRAS, 420, 170

\bibitem[{Leisenring {et~al.}(2012)Leisenring, Skrutskie, Hinz, Skemer, Bailey,
  Eisner, Garnavich, Hoffmann, Jones, Kenworthy, {et~al.}}]{leisenring2012sky}
Leisenring, J., Skrutskie, M., Hinz, P.~M., {et~al.} 2012, in SPIE Proc, 84464F

\bibitem[{Liddle(2007)}]{liddle2007information}
Liddle, A.~R. 2007, MNRASL, 377, L74

\bibitem[{Lindegren {et~al.}(2016)Lindegren, Lammers, Bastian, Hern{\'a}ndez,
  Klioner, Hobbs, Bombrun, Michalik, Ramos-Lerate, Butkevich,
  {et~al.}}]{lindegren2016gaia}
Lindegren, L., Lammers, U., Bastian, U., {et~al.} 2016, A\&A, 595, A4

\bibitem[{Liu(1987)}]{liutipper}
Liu, Z.-Y. 1987, 225 GHz Atmospheric Receiver User's Manual, NRAO Electronics
  Division Internal Report 271

\bibitem[{Longair(2011)}]{longair2011high}
Longair, M.~S. 2011, High Energy Astrophysics (Cambridge University Press)

\bibitem[{Lord(1992)}]{lord1992nasa}
Lord, S. 1992, Ames Research Center, Moffett Field, CA

\bibitem[{Maire {et~al.}(2015)Maire, Skemer, Hinz, Desidera, Esposito, Gratton,
  Marzari, Skrutskie, Biller, Defrere, {et~al.}}]{maire2015leech}
Maire, A.-L., Skemer, A., Hinz, P.~M., {et~al.} 2015, A\&A, 576, A133

\bibitem[{Mandel \& Agol(2002)}]{mandel2002analytic}
Mandel, K., \& Agol, E. 2002, ApJL, 580, L171

\bibitem[{Mandell {et~al.}(2011)Mandell, Deming, Blake, Knutson, Mumma,
  Villanueva, \& Salyk}]{mandell2011non}
Mandell, A.~M., Deming, L.~D., Blake, G.~A., {et~al.} 2011, ApJ, 728, 18

\bibitem[{McCullough {et~al.}(2014)McCullough, Crouzet, Deming, \&
  Madhusudhan}]{mccullough2014water}
McCullough, P., Crouzet, N., Deming, D., \& Madhusudhan, N. 2014, ApJ, 791, 55

\bibitem[{Narita {et~al.}(2013)Narita, Fukui, Ikoma, Hori, Kurosaki, Kawashima,
  Nagayama, Onitsuka, Sukom, Nakajima, {et~al.}}]{narita2013multi}
Narita, N., Fukui, A., Ikoma, M., {et~al.} 2013, ApJ, 773, 144

\bibitem[{Polido {et~al.}(2013)Polido, Jablonski, \& Lepine}]{polido2013galaxy}
Polido, P., Jablonski, F., \& Lepine, J. R.~D. 2013, ApJ, 778, 32

\bibitem[{Polsdofer {et~al.}(2015)Polsdofer, Seale, Sewi{\l}o, Vijh, Meixner,
  Marengo, \& Terrazas}]{polsdofer2015examining}
Polsdofer, E., Seale, J., Sewi{\l}o, M., {et~al.} 2015, AJ, 149, 78

\bibitem[{Pont {et~al.}(2008)Pont, Knutson, Gilliland, Moutou, \&
  Charbonneau}]{pont2008detection}
Pont, F., Knutson, H., Gilliland, R., Moutou, C., \& Charbonneau, D. 2008,
  MNRAS, 385, 109

\bibitem[{Pont {et~al.}(2006)Pont, Zucker, \& Queloz}]{pont2006effect}
Pont, F., Zucker, S., \& Queloz, D. 2006, MNRAS, 373, 231

\bibitem[{Poyneer \& Macintosh(2004)}]{poyneer2004spatially}
Poyneer, L.~A., \& Macintosh, B. 2004, JOSA A, 21, 810

\bibitem[{Prusti {et~al.}(2016)Prusti, De~Bruijne, Brown, Vallenari, Babusiaux,
  Bailer-Jones, Bastian, Biermann, Evans, Eyer, {et~al.}}]{prusti2016gaia}
Prusti, T., De~Bruijne, J., Brown, A., {et~al.} 2016, A\&A, 595, A1

\bibitem[{Quijadaa {et~al.}(2004)Quijadaa, Marxb, Arendtc, \&
  Moseleyb}]{quijadaa2004angle}
Quijadaa, M.~A., Marxb, C.~T., Arendtc, R.~G., \& Moseleyb, S.~H. 2004, in
  Proc. SPIE, Vol. 5487, 245

\bibitem[{Rakich {et~al.}(2011)Rakich, Thompson, \&
  Kuhn}]{rakich2011maintaining}
Rakich, A., Thompson, D., \& Kuhn, O. 2011, in SPIE Proc, 81310H

\bibitem[{Redfield {et~al.}(2008)Redfield, Endl, Cochran, \&
  Koesterke}]{redfield2008sodium}
Redfield, S., Endl, M., Cochran, W.~D., \& Koesterke, L. 2008, ApJL, 673, L87

\bibitem[{Rodigas {et~al.}(2014)Rodigas, Debes, Hinz, Mamajek, Pecaut, Currie,
  Bailey, Defrere, De~Rosa, Hill, {et~al.}}]{rodigas2014does}
Rodigas, T.~J., Debes, J.~H., Hinz, P.~M., {et~al.} 2014, ApJ, 783, 21

\bibitem[{Rodler {et~al.}(2013)Rodler, K{\"u}rster, \&
  Barnes}]{rodler2013detection}
Rodler, F., K{\"u}rster, M., \& Barnes, J.~R. 2013, MNRAS, 432, 1980

\bibitem[{Schwarz {et~al.}(1978)}]{schwarz1978estimating}
Schwarz, G., {et~al.} 1978, The Annals of Statistics, 6, 461

\bibitem[{Simons \& Tokunaga(2002)}]{simons2002mauna}
Simons, D., \& Tokunaga, A. 2002, PASP, 114, 169

\bibitem[{Sing {et~al.}(2009)Sing, D{\'e}sert, Des~Etangs, Ballester,
  Vidal-Madjar, Parmentier, Hebrard, \& Henry}]{sing2009transit}
Sing, D.~K., D{\'e}sert, J.-M., Des~Etangs, A.~L., {et~al.} 2009, A\&A, 505,
  891

\bibitem[{Sing {et~al.}(2011)Sing, D{\'e}sert, Fortney, Des~Etangs, Ballester,
  Cepa, Ehrenreich, L{\'o}pez-Morales, Pont, Shabram, {et~al.}}]{sing2011gran}
Sing, D.~K., D{\'e}sert, J.-M., Fortney, J.~J., {et~al.} 2011, A\&A, 527, A73

\bibitem[{Sing {et~al.}(2012)Sing, Huitson, Lopez-Morales, Pont, D{\'e}sert,
  Ehrenreich, Wilson, Ballester, Fortney, des Etangs, {et~al.}}]{sing2012gtc}
Sing, D.~K., Huitson, C.~M., Lopez-Morales, M., {et~al.} 2012, MNRAS, 426, 1663

\bibitem[{Skemer {et~al.}(2012)Skemer, Hinz, Esposito, Burrows, Leisenring,
  Skrutskie, Desidera, Mesa, Arcidiacono, Mannucci, {et~al.}}]{skemer2012first}
Skemer, A.~J., Hinz, P.~M., Esposito, S., {et~al.} 2012, ApJ, 753, 14

\bibitem[{Skemer {et~al.}(2014)Skemer, Marley, Hinz, Morzinski, Skrutskie,
  Leisenring, Close, Saumon, Bailey, Briguglio, {et~al.}}]{skemer2014directly}
Skemer, A.~J., Marley, M.~S., Hinz, P.~M., {et~al.} 2014, ApJ, 792, 17

\bibitem[{Skrutskie {et~al.}(2006)Skrutskie, Cutri, Stiening, Weinberg,
  Schneider, Carpenter, Beichman, Capps, Chester, Elias,
  {et~al.}}]{skrutskie2006two}
Skrutskie, M., Cutri, R., Stiening, R., {et~al.} 2006, AJ, 131, 1163

\bibitem[{Spalding {et~al.}(2016)Spalding, Skemer, Hinz, \&
  Hill}]{spalding2016infrared}
Spalding, E., Skemer, A., Hinz, P.~M., \& Hill, J.~M. 2016, in SPIE Proc,
  99083C

\bibitem[{Stephens {et~al.}(2001)Stephens, Marley, Noll, \&
  Chanover}]{stephens2001band}
Stephens, D.~C., Marley, M.~S., Noll, K.~S., \& Chanover, N. 2001, ApJL, 556,
  L97

\bibitem[{Swain {et~al.}(2014)Swain, Line, \& Deroo}]{swain2014detection}
Swain, M.~R., Line, M.~R., \& Deroo, P. 2014, ApJ, 784, 133

\bibitem[{Thomas-Osip {et~al.}(2007)Thomas-Osip, McWilliam, Phillips, Morrell,
  Thompson, Folkers, Adams, \& Lopez-Morales}]{thomas2007calibration}
Thomas-Osip, J., McWilliam, A., Phillips, M., {et~al.} 2007, PASP, 119, 697

\bibitem[{Todorov {et~al.}(2009)Todorov, Deming, Harrington, Stevenson, Bowman,
  Nymeyer, Fortney, \& Bakos}]{todorov2009spitzer}
Todorov, K., Deming, D., Harrington, J., {et~al.} 2009, ApJ, 708, 498

\bibitem[{Tokunaga(2000)}]{tokunaga2000allen}
Tokunaga, A. 2000, Allen's Astrophysical Quantities, ed. AN Cox

\bibitem[{Tokunaga {et~al.}(2002)Tokunaga, Simons, \&
  Vacca}]{tokunaga2002mauna}
Tokunaga, A., Simons, D., \& Vacca, W. 2002, PASP, 114, 180

\bibitem[{Torres {et~al.}(2008)Torres, Winn, \& Holman}]{torres2008improved}
Torres, G., Winn, J.~N., \& Holman, M.~J. 2008, ApJ, 677, 1324

\bibitem[{Tyson \& Frazier(2004)}]{tyson2004field}
Tyson, R.~K., \& Frazier, B.~W. 2004, Field guide to adaptive optics, Vol.~2
  (SPIE Press)

\bibitem[{Vrieze(2012)}]{vrieze2012model}
Vrieze, S.~I. 2012, Psychological methods, 17, 228

\bibitem[{Winn(2010)}]{winn2010exoplanet}
Winn, J.~N. 2010, Exoplanets, 1, 55

\bibitem[{Winn {et~al.}(2007)Winn, Holman, Bakos, P{\'a}l, Johnson, Williams,
  Shporer, Mazeh, Fernandez, Latham, {et~al.}}]{winn2007transit}
Winn, J.~N., Holman, M.~J., Bakos, G.~A., {et~al.} 2007, AJ, 134, 1707

\bibitem[{Winn {et~al.}(2008)Winn, Holman, Torres, McCullough, Johns-Krull,
  Latham, Shporer, Mazeh, Garcia-Melendo, Foote, {et~al.}}]{winn2008transit}
Winn, J.~N., Holman, M.~J., Torres, G., {et~al.} 2008, ApJ, 683, 1076

\bibitem[{Zellem {et~al.}(2014)Zellem, Griffith, Deroo, Swain, \&
  Waldmann}]{zellem2014ground}
Zellem, R.~T., Griffith, C.~A., Deroo, P., Swain, M.~R., \& Waldmann, I.~P.
  2014, ApJ, 796, 48

\bibitem[{Zellem {et~al.}(2015)Zellem, Griffith, Pearson, Turner, Henry,
  Williamson, Fitzpatrick, Teske, \& Biddle}]{zellem2015xo}
Zellem, R.~T., Griffith, C.~A., Pearson, K.~A., {et~al.} 2015, ApJ, 810, 11

\end{thebibliography}

\end{document}